\documentclass{article}
\PassOptionsToPackage{numbers,compress}{natbib}
\usepackage[main,final]{neurips_2026}
\usepackage[utf8]{inputenc}
\usepackage[T1]{fontenc}
\usepackage{amsmath,amssymb,amsfonts}
\usepackage{graphicx}
\usepackage[table]{xcolor}
\usepackage{multirow,booktabs}
\usepackage{caption,subcaption}
\usepackage{algorithm,algpseudocode}
\usepackage{hyperref,url}
\usepackage{nicefrac,microtype}
\usepackage{courier}
\usepackage{placeins}
\newcommand{\best}[1]{\textbf{\textcolor{red}{#1}}}
\newcommand{\second}[1]{\textbf{\textcolor{blue}{#1}}}

\hypersetup{pdftitle={BrainVista: Modeling Naturalistic Brain Dynamics as Multimodal Next-Token Prediction},pdfauthor={Xuanhua Yin, Runkai Zhao, Lina Yao, Weidong Cai}}

\title{BrainVista: Modeling Naturalistic Brain Dynamics as Multimodal Next-Token Prediction}

\author{%
  Xuanhua Yin$^{1}$ \quad Runkai Zhao$^{1}$\thanks{Corresponding authors: Runkai Zhao and Weidong Cai.\newline Xuanhua Yin: \texttt{xuanhua.yin@sydney.edu.au}; Lina Yao: \texttt{lina.yao@unsw.edu.au}.} \quad
  Lina Yao$^{2,3}$ \quad Weidong Cai$^{1}$\footnotemark[1] \\[3pt]
  {\normalfont $^{1}$The University of Sydney, Australia} \\
  {\normalfont $^{2}$University of New South Wales, Australia} \\
  {\normalfont $^{3}$CSIRO's Data61} \\[3pt]
  {\normalfont\texttt{runkai.zhao@sydney.edu.au}\quad\texttt{tom.cai@sydney.edu.au}}
}

\begin{document}

\maketitle

\begin{abstract}
Naturalistic fMRI offers a time-resolved window into cortical dynamics during continuous multimodal experience, where future brain states are jointly governed by endogenous neural context and exogenous sensory drive.
Forecasting such dynamics, however, remains challenging because fast-changing stimulus streams must be reconciled with the sluggish hemodynamic BOLD response, while cortical activity is organized across functionally heterogeneous brain networks.
To address these challenges, we introduce BrainVista, a multimodal autoregressive framework for observed-stimulus conditional brain-state forecasting.
BrainVista predicts future fMRI states from the history of brain activity and stimulus tokens temporally aligned to each prediction query, enabling autoregressive prediction without access to future ground-truth fMRI states. Recursive rollout provides a within-run predictive test of brain-state evolution under continuing sensory input.
Specifically, we combine Network-wise Tokenizers for functionally structured brain representations, a Spatial Mixer Head for cross-network interaction refinement, and Stimulus-to-Brain (S2B) masking to preserve temporally causal stimulus--brain conditioning while preventing future brain-state leakage and stimulus look-ahead.
We evaluate BrainVista on Algonauts 2025, CineBrain, and HAD, three large-scale naturalistic fMRI benchmarks, under long-horizon conditional autoregressive forecasting.
BrainVista consistently outperforms baselines, improving pattern correlation by 10.8\% and 9.4\% relative to the strongest baseline on Algonauts 2025 and CineBrain, respectively.
Code is available at \url{https://github.com/xuanhuayin/BrainVista}.
\end{abstract}

% \newpage
\section{Introduction}

The Blood-Oxygen-Level-Dependent (BOLD) signal in functional magnetic resonance imaging (fMRI) is a temporally structured measurement of brain activity, where future observations are constrained by prior neural and hemodynamic states rather than arising as independent samples \citep{granger1969,geweke1982,logothetis2001neurophysiological}.
Beyond short-range autocorrelation, this temporal structure reflects the brain's continuous integration of intrinsic dynamics with incoming sensory evidence, a view closely related to predictive accounts of cortical processing \citep{friston2005theory,clark2013whatever}.
This coupling is particularly salient during naturalistic stimulation, such as movie viewing, where cortical responses evolve under continuous visual, auditory, and semantic input.
We therefore view naturalistic fMRI as an observed, non-autonomous dynamical process: the future brain state depends on prior brain activity and an exogenous stimulus stream that is time-aligned to the fMRI acquisition grid.

The scientific question is whether a model can sustain the complementary information in brain history and observed stimuli when its own predictions replace measured brain states. We use within-run forecasting as a predictive assay: observed fMRI initializes the state, and the withheld continuation tests competing dynamical assumptions under identical input constraints. Input ablations test the contribution of each source; history occlusion tests predictive reliance on specific lags; and long-horizon rollout distinguishes one-step fitting from stable recursive prediction. This scope does not require treating forecasts as faithful brain simulations.

Despite the temporal nature of BOLD responses, many recent brain foundation models are trained with masked reconstruction, bidirectional attention, or global-context representation learning objectives \citep{caro2024brainlm,dong2024brainjepa,dong2025brainharmony}.
These objectives are effective for self-supervised representation learning and downstream transfer, but they are not directly designed for autoregressive forecasting of future brain states.
When bidirectional or masked-context models are used for forecasting, they may create a training--inference mismatch: during training, the model can exploit information from time points that would be unavailable during recursive rollout.
This mismatch is especially problematic for long-horizon prediction, where future ground-truth fMRI states must be withheld and model predictions are recursively fed back as context.
Thus, naturalistic fMRI forecasting requires an evaluation protocol and model architecture that distinguish observed exogenous stimulus inputs from unavailable future brain states.

Autoregressive models and other sequence architectures have advanced language modeling and time-series forecasting \citep{liu2025timerxl,liu2024itransformer,wu2023timesnet,nie2022patchtst,brown2020gpt3}.
However, extending this paradigm to naturalistic fMRI is non-trivial.
First, multimodal stimulus streams are sampled at substantially higher temporal resolution than fMRI, while BOLD responses are slow and hemodynamically delayed \citep{lahner2024boldmoments,chang2022timescalehierarchy}.
A forecasting model must therefore align high-frequency sensory features to the fMRI grid without introducing look-ahead from stimulus tokens beyond the queried prediction step.
In our setting the stimulus stream is treated as an observed exogenous driver: predicting the brain state at time $t{+}1$ may condition on the stimulus token aligned to that target time, but not on future stimulus tokens beyond $t{+}1$ or on the target and future fMRI states.
Second, cortical responses are organized across heterogeneous functional systems.
The cortex contains specialized networks with conserved functional organization \citep{glasser2016multimodal,finn2015fingerprinting,gratton2018stable}, and hierarchical differences in circuit and temporal properties shape large-scale dynamics \citep{demirtas2019hierarchical,wang2025foundationmodel,bayazi2024bfm}.
Generic sequence models that compress the whole brain into a single homogeneous representation can blur functional boundaries, reduce network-specific contrast, and obscure structured interactions across cortical systems.

To address these challenges, we propose BrainVista, a time-aligned multimodal framework that combines network-specific representations, directed stimulus--brain attention, and cross-network refinement for recursive fMRI forecasting.
Our contributions are threefold:
\begin{itemize}
\item We formulate naturalistic fMRI modeling as observed-stimulus conditional autoregressive forecasting and develop BrainVista to predict future brain states from brain history and available stimuli. Its Stimulus-to-Brain (S2B) mask enforces temporally valid conditioning while keeping stimulus representations independent of brain-token history.

\item We combine Network-wise Tokenizers with a Spatial Mixer Head to retain separate network representations before temporal prediction and refine predicted network components before decoding. Controlled ablations evaluate these design choices and their interaction during recursive forecasting.

\item Experiments on Algonauts 2025~\citep{gifford2025algonauts}, CineBrain~\citep{gao2025cinebrain}, and HAD~\citep{zhou2023had} show consistent long-horizon improvements over matched-access baselines, including Transformer-ARX with the same frozen encoders and tokenizers. Additional experiments assess tokenizer objectives, data and model scaling, and the representational structure retained by forecasts.
\end{itemize}

\section{Related Work}

\textbf{fMRI Tokenization and Brain Foundation Models.}
High-dimensional fMRI signals require compact representations for scalable sequence modeling.
Existing approaches include discrete code sequences \citep{nishimura2024braincodec,yang2025hst}, task-aware reconstruction \citep{zhao2024tardrl}, probabilistic latent tokens \citep{mai2025synbrain,li2025brainvae}, and masked-reconstruction pretraining \citep{qu2025uncovering,xia2026tlmae}.
Recent brain foundation models learn transferable neural representations from large-scale neuroimaging data, including Brain-JEPA \citep{dong2024brainjepa}, Brain Harmony \citep{dong2025brainharmony}, and BrainLM \citep{caro2024brainlm}, with related extensions to atlas-free, memory-efficient, concept-conditioned, and graph-based modeling \citep{wang2025slimbrain,bao2025mindsimulator,wei2025braingfm}.
These methods mainly target reconstruction, representation learning, or task transfer, whereas BrainVista focuses on temporally constrained forecasting under observed naturalistic stimuli.

\textbf{Multimodal Stimulus--Brain Modeling.}
Naturalistic movie paradigms motivate multimodal modeling of brain activity from temporally aligned visual, auditory, and language inputs.
TRIBE predicts whole-brain responses from video, audio, and text features~\citep{tribe2025}; SynBrain models one-to-many stimulus-to-fMRI mappings~\citep{mai2025synbrain}; dynamic subject-aware routing addresses inter-subject variability~\citep{yin2025dynamicrouting}; and Algonauts 2025 provides a large-scale benchmark for naturalistic fMRI prediction~\citep{gifford2025algonauts}.
Related decoding studies reconstruct stimuli from brain activity using diffusion-based or unified multimodal models~\citep{scotti2024mindeye2,xia2024umbrae,takagi2023ldmfmri,scotti2023mindeye}.
Bratrix explores language-anchored vision--brain alignment with uncertainty-aware weighting~\citep{feng2025unveilingdeepsemanticuncertainty}.
In contrast, BrainVista predicts future brain states from past brain activity and aligned sensory context, treating naturalistic fMRI as an evolving state process.

\textbf{Autoregressive Brain-State Forecasting.}
Autoregressive modeling is well suited to temporally evolving brain activity, and prior work has studied individual fMRI time-series forecasting in large-scale datasets~\citep{paugam2024autoreg}.
This direction is also supported by evidence for hierarchical temporal receptive windows and heterogeneous intrinsic timescales during naturalistic cognition~\citep{hasson2008hierarchy,lerner2011topographic,honey2012slow,murray2014hierarchy,chang2022timescalehierarchy}.
However, naturalistic fMRI forecasting must combine endogenous brain history with exogenous multimodal stimuli while avoiding leakage from mismatched stimulus and BOLD timescales~\citep{lahner2024boldmoments,chang2022timescalehierarchy}.
BrainVista addresses this setting through Stimulus-to-Brain masking, Network-wise Tokenizers, and Spatial Mixer-based cross-network refinement.

\section{Methodology}
\label{sec:method}

\begin{figure*}[!t]
  \centering
  \includegraphics[width=\linewidth]{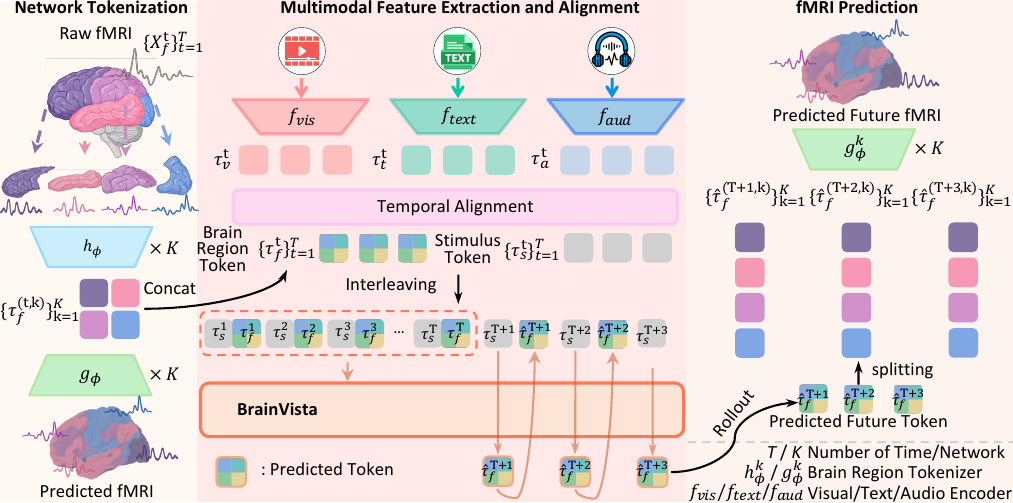}
\caption{\textbf{Framework of BrainVista.}
Time-aligned multimodal features are aggregated into stimulus tokens, while parcel-wise fMRI is encoded by network-specific tokenizers into functionally structured brain tokens.
Given interleaved stimulus and brain tokens from the valid context, BrainVista predicts the next brain token and decodes it back to fMRI space.
}
  \label{fig:pipeline}
  \vskip -0.15in
\end{figure*}

BrainVista forecasts naturalistic fMRI as time-aligned multimodal next-token prediction without access to target or future fMRI.
As shown in Fig.~\ref{fig:pipeline}, it aligns observed stimulus features to the fMRI grid, encodes network-wise brain tokens, and predicts future tokens with a Transformer under Stimulus-to-Brain masking.
The Spatial Mixer Head refines cross-network interactions before decoding to fMRI space.

\subsection{Problem Setup and Notation}
\label{subsec:problem}

\textbf{Observed-stimulus Conditional Brain Forecasting.}
For each subject and run, let $x^t\in\mathbb{R}^{P}$ denote the model-space fMRI state at time step $t$, where $P$ is the number of parcels or ROI features.
Let $\tau_s^t$ denote the stimulus token aligned to the same fMRI grid.
The stimulus stream is treated as an observed external driver, whereas future fMRI states are latent variables to be predicted.

For a queried prediction time $u$, the valid conditioning set is:
\begin{equation}
\mathcal{C}_u
=
\left(
x^{<u},\tau_s^{\leq u}
\right),
\label{eq:valid_conditioning}
\end{equation}
where the target-aligned stimulus token $\tau_s^u$ is observed, but the target and future fMRI states $x^{\geq u}$ are never used as inputs.

\textbf{One-step Transition.}
At time $t$, BrainVista learns a one-step transition:
\begin{equation}
\hat{x}^{t+1}
=
f_{\theta}
\left(
x^{\leq t},
\tau_s^{\leq t+1}
\right),
\label{eq:one_step_transition}
\end{equation}
where $f_{\theta}$ is the forecasting model,
$x^{\leq t}=(x^1,\ldots,x^t)$ denotes the ordered brain-state history, and
$\tau_s^{\leq t+1}=(\tau_s^1,\ldots,\tau_s^{t+1})$ denotes the ordered stimulus sequence up to the queried prediction step.

\textbf{Multi-step Conditional Rollout.}
Given an observed context window ending at time $T$, an $H$-step forecast is produced recursively:
\begin{equation}
\hat{x}^{T+h}
=
f_{\theta}
\left(
x^{\leq T};
\hat{x}^{T+1:T+h-1},
\tau_s^{\leq T+h}
\right),
\qquad
h=1,\ldots,H,
\label{eq:conditional_rollout}
\end{equation}
where $H$ is the rollout horizon and
$\hat{x}^{T+1:T+h-1}=(\hat{x}^{T+1},\ldots,\hat{x}^{T+h-1})$
denotes the ordered sequence of previously predicted brain states, which is empty when $h=1$.
The stimulus sequence $\tau_s^{\leq T+h}$ remains observed up to the queried prediction step.
Thus, our setting is conditional brain-state forecasting under observed naturalistic stimuli; it neither models unconditional neural dynamics nor establishes biological causality.

\subsection{Stimulus Tokens}
\label{subsec:stimuli}

\textbf{Frozen Encoders.}
We employ fixed pretrained encoders $f_{\mathrm{vis}}, f_{\mathrm{aud}}, f_{\mathrm{text}}$ to map raw video, audio, and text streams $s_{\mathrm{vis}}, s_{\mathrm{aud}}, s_{\mathrm{text}}$ into time-indexed stimulus features, producing $\tau_{\mathrm{vis}}\in\mathbb{R}^{d_{\mathrm{v}}}$, $\tau_{\mathrm{aud}}\in\mathbb{R}^{d_{\mathrm{a}}}$, and $\tau_{\mathrm{text}}\in\mathbb{R}^{d_{\mathrm{t}}}$.
Here, $d_{\mathrm{v}}$, $d_{\mathrm{a}}$, and $d_{\mathrm{t}}$ denote the embedding dimensionalities of the video, audio, and text stimulus features, respectively.

\textbf{Temporal Alignment.}
We aggregate stimulus features onto the native fMRI grid, treating each fMRI timestamp as the end of an acquisition bin.
Video and audio features are eligible only if their extraction windows end no later than that bin boundary; text uses only the causal prefix available by the same boundary.
For modality $m\in\{\mathrm{vis},\mathrm{aud},\mathrm{text}\}$, let $\tau'_{m,i}\in\mathbb{R}^{d_m}$ denote the $i$-th high-rate feature, and let $\mathcal{B}_{m,t}$ denote the set of features assigned to fMRI bin $t$.
We mean-pool the eligible features assigned to each bin:
\begin{equation}
\tau_m^t
=
\frac{1}{|\mathcal{B}_{m,t}|}
\sum_{i\in\mathcal{B}_{m,t}}
\tau'_{m,i},
\qquad
t=1,\ldots,T .
\label{eq:stim_bin_pooling}
\end{equation}

\textbf{Shared Stimulus Token.}
At each fMRI time step $t$, we form a shared stimulus token by concatenating modality tokens and projecting them to the Transformer hidden space:
\begin{equation}
\tau_s^t
=
\mathrm{Proj}_s
\left(
[
\tau_{\mathrm{vis}}^t;
\tau_{\mathrm{aud}}^t;
\tau_{\mathrm{text}}^t
]
\right)
\in \mathbb{R}^{d_{\mathrm{model}}},
\label{eq:shared_stim_token}
\end{equation}
where $\mathrm{Proj}_s(\cdot)$ is a learned projection and $d_{\mathrm{model}}$ denotes the token embedding dimensionality of the temporal Transformer.
We add no separate video, audio, or text modality embeddings because these features form one stimulus token.
Dataset-specific TRs, extraction windows, strides, and modality availability are detailed in Appendix~\ref{sec:appx_temporal_alignment} and Table~\ref{tab:timestamp_feature_windows}.

\subsection{Brain Region Tokens}
\label{subsec:tokenizer}

\textbf{Network-aware Grouping.}
Because cortical systems exhibit heterogeneous temporal and functional properties~\citep{yeo2011organization,glasser2016multimodal}, we partition the $P$ fMRI dimensions into $K$ functional networks $\{ \mathcal{V}_k \}_{k=1}^{K}$.
This avoids compressing the whole brain into a single homogeneous representation, which can over-smooth network-specific structure~\citep{qiu2025mindllm}.
Each $\mathcal{V}_k$ is a fixed, nonlearned index set: Algonauts 2025 and HAD use Schaefer-to-Yeo-7 assignments~\citep{yeo2011organization,schaefer2018parcellation}, whereas CineBrain uses dataset-provided visual and auditory ROI groups.
The selected signals remain multidimensional vectors, without averaging within groups.
Given the native fMRI response $x_{\mathrm{raw}}^t$, the model-space signal of group $k$ is:
\begin{equation}
x^{k,t}
=
\mathcal{A}_k\!\left(x_{\mathrm{raw}}^t[\mathcal{V}_k]\right)
\in\mathbb{R}^{P_k},
\label{eq:network_index}
\end{equation}
where $\mathcal{A}_k$ is identity mapping for parcellated inputs and a training-set PCA projection for CineBrain.
All preprocessing statistics are fitted on training runs only and fixed for validation, test, and rollout evaluation.
The whole-brain representation is
$x^t=[x^{1,t};\ldots;x^{K,t}]\in\mathbb{R}^{P}$, where $P=\sum_{k=1}^{K}P_k$.

\textbf{Network-wise Tokenization.}
We use independent network-wise tokenizer--detokenizer pairs to obtain compact brain tokens.
For group $k$, a lightweight MLP tokenizer $h_{\phi_k}^k$ and detokenizer $g_{\psi_k}^k$ compute:
\begin{equation}
\tau_f^{t,k}
=
h_{\phi_k}^k(x^{k,t})
\in\mathbb{R}^{d_k},
\qquad
\tilde{x}^{k,t}
=
g_{\psi_k}^k(\tau_f^{t,k})
\in\mathbb{R}^{P_k},
\label{eq:tokenize_detokenize}
\end{equation}
where $\tau_f^{t,k}$ is the brain region token, $\tilde{x}^{k,t}$ is the reconstructed model-space signal, and $d_k$ is the network-specific token dimensionality.
The parameters $\phi_k$ and $\psi_k$ are network-specific and are not shared across functional networks.
Dataset-specific values of $P_k$, $d_k$, and $d_{\mathrm{brain}}$ are reported in Appendix~\ref{sec:appx_hyperparameters}.

\textbf{Projection to Transformer Space.}
The group-wise brain region tokens are concatenated into a whole-brain token and projected to the Transformer hidden space before temporal modeling:
\begin{equation}
\tau_f^t
=
[\tau_f^{t,1};\ldots;\tau_f^{t,K}]
\in\mathbb{R}^{d_{\mathrm{brain}}}.
\label{eq:brain_token_concat}
\end{equation}
A learned input projection maps this whole-brain token to the same Transformer token dimension as the stimulus token.
For notational simplicity, we continue to denote the projected brain token by $\tau_f^t$ in the temporal Transformer.

\begin{figure}[!t]
  \centering
  \includegraphics[width=0.9\linewidth]{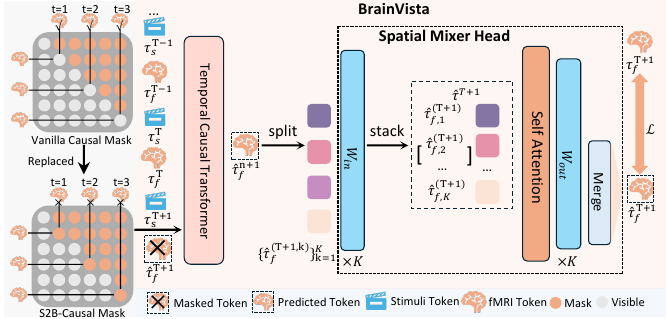}
\caption{\textbf{BrainVista core.}
For next-step prediction, BrainVista takes an interleaved sequence of stimulus tokens and available brain tokens from the recent context.
The target brain-token position is decoded as the prediction target.
The Stimulus-to-Brain mask keeps stimulus tokens in a stimulus-only causal stream.
The temporal Transformer predicts a whole-brain latent token, which is refined by the Spatial Mixer Head and decoded to fMRI space.}
  \label{fig:arch}
  \vskip -0.15in
\end{figure}

\subsection{BrainVista: Temporal Causal Transformer with Stimulus-to-Brain Mask}
\label{subsec:BrainVista}

\textbf{Interleaved Time-aligned Input Sequence.}
For a queried prediction time $u$, we interleave the recent brain and stimulus windows, whose exact index ranges are given in Appendix~\ref{sec:appx_dataset_statistics}.
The stimulus token at the queried time is available, while the corresponding brain-token position is decoded as the prediction target.
Fig.~\ref{fig:arch} illustrates this core computation, including the interleaved input sequence, the S2B-causal attention mask, target-token prediction, and Spatial Mixer refinement.
Writing $N_u$ for the sequence length, the ordered projected tokens have the form:
\begin{equation}
S_u
=
\big(
\ldots,
\tau_s^{u-1}, \tau_f^{u-1},
\tau_s^{u}, \tau_f^{u}
\big)
\in
\mathbb{R}^{N_u\times d_{\mathrm{model}}}.
\label{eq:interleave}
\end{equation}
Here, the final brain-token position $\tau_f^{u}$ denotes the target position to be decoded, rather than an observed brain token encoded from $x^u$.
This construction preserves the target-aligned stimulus token $\tau_s^u$ while preventing access to the unavailable brain state $x^u$.
Separate stimulus and brain input projections, fixed interleaved positions, positional embeddings, and S2B attention rules distinguish the two token types.

\textbf{Stimulus-to-Brain Mask (S2B).}
We constrain self-attention with a Stimulus-to-Brain mask $M_u\in\{0,1\}^{N_u\times N_u}$ over $S_u$.
Stimulus-token queries attend only to non-future stimulus tokens, forming a stimulus-only causal stream.
Brain-token queries attend to causally available stimulus tokens and past brain tokens.
The target brain-token position $\tau_f^u$ attends to $\tau_s^{\leq u}$ and $\tau_f^{<u}$, but never to the target or future fMRI states.
This preserves the observed-stimulus conditional rollout graph without stimulus look-ahead beyond the queried step.
Unlike S2B, vanilla causal attention allows stimulus queries to read earlier brain tokens, including predictions during rollout, so prediction errors can enter stimulus representations.
S2B blocks this brain-to-stimulus path; its defining distinction is directed information flow, beyond target-aligned stimulus availability.

We feed $S_u$ into a GPT-style Transformer under $M_u$, take the hidden state at the target brain-token position, and apply a brain output projection to predict the preliminary whole-brain token:
\begin{equation}
H_u
=
\mathrm{Transformer}(S_u;M_u),
\qquad
\tilde{\tau}_f^{u}
=
\mathrm{Proj}_{b}^{\mathrm{out}}\!\left(H_u[\tau_f^u]\right)
\in\mathbb{R}^{d_{\mathrm{brain}}},
\label{eq:transformer_pred}
\end{equation}
where $H_u\in\mathbb{R}^{N_u\times d_{\mathrm{model}}}$ is the sequence of hidden states,
$H_u[\tau_f^u]$ denotes the hidden state at the target brain-token position, and
$\tilde{\tau}_f^{u}$ is the preliminary whole-brain latent prediction.

\textbf{Spatial Mixer Head.}
To refine cross-network interactions, we split $\tilde{\tau}_f^{u}$ into network-wise components $\{\tilde{\tau}_f^{u,k}\}_{k=1}^{K}$ and project them to a shared mixer dimension:
\begin{equation}
R^{u}
=
\mathrm{Stack}\!\left(
W_{\mathrm{in}}^{1}\tilde{\tau}_f^{u,1},
\ldots,
W_{\mathrm{in}}^{K}\tilde{\tau}_f^{u,K}
\right)
\in\mathbb{R}^{K\times d_{\mathrm{mix}}}.
\label{eq:mixer_stack}
\end{equation}
A multi-head self-attention layer models state-dependent interactions across the $K$ network tokens:
\begin{equation}
\tilde{R}^{u}
=
\mathrm{MHA}(R^{u}).
\label{eq:mixer_attn}
\end{equation}
The outputs are mapped back to network-wise token dimensions and concatenated:
\begin{equation}
\hat{\tau}_f^{u,k}
=
W_{\mathrm{out}}^{k}\tilde{R}_{k}^{u},
\qquad
\hat{\tau}_f^{u}
=
\big[
\hat{\tau}_f^{u,1};
\dots;
\hat{\tau}_f^{u,K}
\big]
\in\mathbb{R}^{d_{\mathrm{brain}}}.
\label{eq:concat_refined}
\end{equation}
The attention matrix provides a diagnostic summary of pairwise network coupling, but is not interpreted as biological causality.

\textbf{Decoding to fMRI Space.}
The refined network-wise tokens are decoded by the frozen detokenizers:
\begin{equation}
\hat{x}^{u,k}
=
g_{\psi_k}^{k}(\hat{\tau}_f^{u,k}),
\qquad
\hat{x}^{u}
=
\operatorname{Concat}_{k=1}^{K}
\hat{x}^{u,k},
\label{eq:decode_pred}
\end{equation}
where $\hat{x}^{u,k}$ is the predicted model-space signal for network $k$ and $\hat{x}^{u}$ is the predicted whole-brain fMRI state.
When metrics are computed in the original signal scale, predictions are mapped back using the corresponding training-split preprocessing statistics.

\subsection{Training Objectives and Optimization}
\label{subsec:training}

\textbf{Tokenizer Reconstruction.}
We pretrain deterministic network-wise autoencoders using reconstruction alone, with no additional latent prior.
Their low-dimensional bottlenecks, network-specific parameters, full-signal reconstruction, and subsequent freezing constrain the learned representations; the tokens are not random projections.
Appendix~\ref{sec:appx_tokenizer_regularization} compares alternative latent regularizers.
The tokenizer loss is:
\begin{equation}
\mathcal{L}_{\mathrm{tok}}
=
\sum_{t}\sum_{k=1}^{K}
\left\|
x^{k,t}
-
g^{k}_{\psi_k}\!\Big(h^{k}_{\phi_k}(x^{k,t})\Big)
\right\|_2^2 .
\label{eq:loss_tok}
\end{equation}
Here, $\phi_k$ and $\psi_k$ denote the network-specific tokenizer and detokenizer parameters, respectively.

\textbf{Next-token Prediction.}
After tokenizer pretraining, we freeze the tokenizer--detokenizer pairs
$\{h_{\phi_k}^k,g_{\psi_k}^k\}_{k=1}^{K}$ and train BrainVista with teacher forcing to predict the next whole-brain token.
Let $\tau_f^{t+1}\in\mathbb{R}^{d_{\mathrm{brain}}}$ denote the target whole-brain token at time $t{+}1$, and let $\hat{\tau}_f^{t+1}\in\mathbb{R}^{d_{\mathrm{brain}}}$ be the model prediction after the Spatial Mixer Head.
We minimize:
\begin{equation}
\mathcal{L}_{\mathrm{pred}}
=
\sum_{t}
\left\|
\tau_f^{t+1}
-
\hat{\tau}_f^{t+1}
\right\|_2^2 .
\label{eq:loss_pred}
\end{equation}

\section{Experiments and Results}
\label{sec:experiments}

\begin{table}[!t]
\centering
\setlength{\tabcolsep}{3.2pt}
\renewcommand{\arraystretch}{1.05}
\small
\caption{
\textbf{Main results for conditional autoregressive forecasting with horizon $H{=}20$ on Algonauts~2025, CineBrain, and HAD.}
Non-ARX references are reported as references, while ARX baselines receive matched past fMRI states and aligned stimulus features.
Entries are means; seed-level and subject-level statistics are reported in Tables~\ref{tab:seed_variance} and~\ref{tab:subject_level_stability}. Higher is better. Best and second-best means are shown in \textcolor{red}{red} and \textcolor{blue}{blue}.
}
\label{tab:main_results_rpcorr_cls}
\resizebox{\textwidth}{!}{%
\begin{tabular}{@{}lccc|ccc|ccc@{}}
\toprule
& \multicolumn{3}{c|}{Algonauts~2025}
& \multicolumn{3}{c|}{CineBrain}
& \multicolumn{3}{c}{HAD} \\
\cmidrule(lr){2-4}\cmidrule(lr){5-7}\cmidrule(lr){8-10}
Model
& $r$ & $p_{\mathrm{corr}}$ & Cls@Top10
& $r$ & $p_{\mathrm{corr}}$ & Cls@Top10
& $r$ & $p_{\mathrm{corr}}$ & Cls@Top10 \\
\midrule
\multicolumn{10}{@{}l}{\textit{Non-ARX references}} \\

TRIBE~\citep{tribe2025}
& 0.1379 & 0.1323 & 0.6823
& 0.4742 & 0.3012 & 0.8560
& 0.2685 & 0.1845 & 0.7012 \\

MindSimulator~\citep{bao2025mindsimulator}
& 0.1432 & 0.1478 & 0.6919
& 0.4856 & 0.3235 & 0.8615
& 0.2610 & 0.1960 & 0.7105 \\

BrainLM~\citep{caro2024brainlm}
& 0.1191 & 0.0829 & 0.6493
& 0.4633 & 0.2762 & 0.8478
& 0.2465 & 0.1450 & 0.6912 \\

BrainJEPA~\citep{dong2024brainjepa}
& 0.1493 & 0.0916 & 0.6812
& 0.4941 & 0.2445 & 0.8455
& 0.2515 & 0.2353 & 0.7321 \\

Brain Harmony~\citep{dong2025brainharmony}
& 0.1867 & 0.1466 & 0.7031
& 0.5397 & 0.3484 & 0.8642
& 0.2642 & 0.2688 & 0.7685 \\

\midrule
\multicolumn{10}{@{}l}{\textit{ARX baselines}} \\

Ridge-ARX
& 0.1762 & 0.1584 & 0.7106
& 0.5483 & 0.4016 & 0.8678
& 0.2630 & 0.2327 & 0.7446 \\

LSTM-ARX
& 0.1905 & 0.1718 & 0.7223
& 0.5614 & 0.4529 & 0.8735
& 0.2642 & 0.2605 & 0.7631 \\

TemporalCNN-ARX
& 0.1968 & 0.1756 & 0.7268
& 0.5732 & 0.4724 & 0.8786
& 0.2654 & 0.2741 & 0.7762 \\

Transformer-ARX
& 0.2026 & 0.1819 & 0.7315
& 0.5897 & 0.4978 & 0.8837
& 0.2678 & 0.2724 & 0.7836 \\

BrainHarmony-AR
& \second{0.2065} & \second{0.1887} & \second{0.7392}
& \second{0.5988} & \second{0.5156} & \second{0.8893}
& \second{0.2685} & \second{0.2769} & \second{0.7894} \\

\midrule
{\bfseries BrainVista}
& \best{0.2183} & \best{0.2090} & \best{0.7619}
& \best{0.6315} & \best{0.5643} & \best{0.9012}
& \best{0.2739} & \best{0.2950} & \best{0.7998} \\

\bottomrule
\end{tabular}%
}
\end{table}

\begin{table}[!t]
\centering
\setlength{\tabcolsep}{4.4pt}
\renewcommand{\arraystretch}{1.08}
\small
\caption{
\textbf{HAD forecasting results across horizons for matched-access autoregressive methods.}
We report $H\in\{1,10,20\}$.
Best and second-best values are shown in \textcolor{red}{red} and \textcolor{blue}{blue}.
}
\label{tab:had_forecasting_horizons}
\begin{tabular}{@{}lcc|cc|cc@{}}
\toprule
& \multicolumn{2}{c|}{$H{=}1$}
& \multicolumn{2}{c|}{$H{=}10$}
& \multicolumn{2}{c}{$H{=}20$} \\
\cmidrule(lr){2-3}\cmidrule(lr){4-5}\cmidrule(lr){6-7}
\textbf{Model}
& $r$ & $p_{\mathrm{corr}}$
& $r$ & $p_{\mathrm{corr}}$
& $r$ & $p_{\mathrm{corr}}$ \\
\midrule
Ridge-ARX
& 0.5740 & 0.5530
& 0.2860 & 0.2675
& 0.2630 & 0.2327 \\

LSTM-ARX
& 0.5815 & 0.5740
& 0.3002 & 0.2940
& 0.2642 & 0.2605 \\

TemporalCNN-ARX
& 0.5869 & 0.5831
& 0.3065 & 0.3072
& 0.2654 & 0.2741 \\

Transformer-ARX
& 0.5890 & 0.5876
& 0.3112 & 0.3160
& 0.2678 & 0.2724 \\

BrainHarmony-AR
& \second{0.5912} & \second{0.5925}
& \second{0.3136} & \second{0.3215}
& \second{0.2685} & \second{0.2769} \\

\textbf{BrainVista}
& \best{0.6058} & \best{0.6214}
& \best{0.3278} & \best{0.3390}
& \best{0.2739} & \best{0.2950} \\
\bottomrule
\end{tabular}
\vskip -15pt
\end{table}

\subsection{Experimental Setup}
\label{subsec:exp_setup}

\textbf{Datasets and Splits.}
We evaluate BrainVista on Algonauts~2025~\citep{gifford2025algonauts}, CineBrain~\citep{gao2025cinebrain}, and HAD~\citep{zhou2023had}, covering diverse stimulus paradigms, temporal resolutions, preprocessing pipelines, and brain-space definitions.
For each dataset, we follow the official run-level train/validation/test split, use validation runs only for model selection, and report final results on held-out test runs.
Forecasting windows are constructed within runs and never cross run boundaries.
Algonauts~2025 and HAD use Schaefer-1000 parcels mapped to Yeo-7 systems, while CineBrain uses dataset-defined visual and auditory ROI groups with training-set PCA reduction before tokenization.

\textbf{Brain and Stimulus Representation.}
The fMRI signal is represented on each dataset's native parcel, voxel, or ROI grid and grouped into functional systems for network-wise tokenization.
Algonauts~2025 and HAD use Schaefer-to-Yeo assignments~\citep{schaefer2018parcellation,yeo2011organization}, while CineBrain follows its dataset-defined sensory systems~\citep{gao2025cinebrain}.
All normalization statistics are estimated from training runs only and fixed for validation, test, and rollout evaluation.
When available, video, audio, and text features are extracted using frozen V-JEPA~2~\citep{assran2025vjepa2}, Wav2Vec2-BERT~\citep{seamless2023}, and Llama~3.2~3B~\citep{llamateam2024llama3herd,meta2024llama32}, average-pooled within each fMRI bin, and projected to stimulus tokens.

\textbf{Baselines.}
We compare with non-ARX references from stimulus-to-brain prediction and brain representation learning, including TRIBE~\citep{tribe2025}, MindSimulator~\citep{bao2025mindsimulator}, BrainLM~\citep{caro2024brainlm}, BrainJEPA~\citep{dong2024brainjepa}, and Brain Harmony~\citep{dong2025brainharmony}.
We further evaluate matched-access autoregressive baselines with exogenous inputs (ARX): Ridge-ARX, LSTM-ARX, TemporalCNN-ARX, Transformer-ARX, and BrainHarmony-AR.
Here, ARX denotes models that predict future fMRI states from both endogenous brain history and exogenous observed stimulus tokens, while differing in the forecasting backbone.
BrainHarmony-AR denotes an autoregressive adaptation of Brain Harmony, where its brain representations are used under the same causal forecasting protocol.
All ARX baselines use past fMRI and observed stimuli through the queried bin, with target and future ground-truth fMRI withheld. Transformer-ARX additionally shares BrainVista's frozen stimulus encoders, pretrained Network-wise Tokenizer--detokenizer pairs, four-layer/256-hidden temporal backbone size, splits, context, optimization protocol, and tuning budget (Appendix~\ref{sec:appx_baseline_capacity}). Non-ARX references have different input access: TRIBE is stimulus-only, whereas BrainLM and BrainJEPA use past fMRI (Table~\ref{tab:baseline_input_access}).

\textbf{Training and Metrics.}
Unless otherwise specified, models are trained for 40 epochs with AdamW, mixed precision, learning rate $\eta=2\times10^{-4}$, weight decay $10^{-2}$, and batch size 32.
The default context length and rollout horizon are $L=40$ and $H=20$, respectively.
At test time, models are initialized with observed fMRI context and rolled out autoregressively while the stimulus stream remains observed up to each queried prediction step.
We report vectorized Pearson correlation $r$, pattern correlation $p_{\mathrm{corr}}$, and Top-10 Identification Accuracy (Cls@Top10).
Additional protocol and baseline details are provided in Appendices~\ref{sec:appx_protocol} and~\ref{sec:appx_baseline_adaptation}. Model size, memory usage, inference time, and additional compute details are reported in Appendix~\ref{sec:appx_complexity}.

\subsection{Main Results}
\label{subsec:main_results}

All methods are evaluated under the same conditional autoregressive forecasting protocol when applicable, and all results are averaged over subjects.
For models without a native forecasting interface, we train a forecasting head under the same input-access constraints, with details in Appendix~\ref{sec:appx_baseline_adaptation}.

Table~\ref{tab:main_results_rpcorr_cls} shows that BrainVista performs best on all three datasets under $H{=}20$ forecasting.
Non-ARX references provide useful references, while ARX baselines give stronger matched-access comparisons because they use past fMRI states and aligned stimulus features.
BrainHarmony-AR is the strongest non-BrainVista baseline. BrainVista also exceeds the representation-matched Transformer-ARX by $0.0271$, $0.0665$, and $0.0226$ in $p_{\mathrm{corr}}$ on Algonauts, CineBrain, and HAD, respectively. Thus, identical stimulus encoders and tokenizers do not eliminate its advantage.
The improvements are clearest in $p_{\mathrm{corr}}$, suggesting that BrainVista better preserves instantaneous cross-region brain-state structure rather than only improving global temporal similarity.
Across the three datasets, the standard deviation of $p_{\mathrm{corr}}$ over five random seeds ranges from $0.0022$ to $0.0042$ for BrainVista and from $0.0026$ to $0.0048$ for BrainHarmony-AR (Table~\ref{tab:seed_variance}). Separately, paired subject-bootstrap 95\% confidence intervals for their mean $p_{\mathrm{corr}}$ difference are $[0.0172,0.0234]$, $[0.0449,0.0523]$, and $[0.0171,0.0191]$ on Algonauts, CineBrain, and HAD, with gains for $4/4$, $6/6$, and $30/30$ subjects, respectively (Table~\ref{tab:subject_level_stability}; Appendix~\ref{sec:appx_statistical_reliability}). Algonauts has only four subjects: despite its positive bootstrap interval, the one-sided exact sign test has $p=0.0625$, limiting population-level inference.

Table~\ref{tab:had_forecasting_horizons} reports HAD results across forecasting horizons for matched-access autoregressive methods.
All methods degrade as $H$ increases, reflecting accumulated rollout error under recursive prediction.
BrainVista remains consistently strongest, with a more pronounced margin in $p_{\mathrm{corr}}$ than in vectorized $r$, suggesting that its network-wise tokenization and cross-network refinement better preserve spatial brain-state structure over long horizons.
The advantage persists across the evaluated horizons; component ablations below examine how individual effects change with rollout length. Non-ARX stimulus-to-brain references are summarized in Appendix~\ref{sec:appx_full_horizon_native_refs}, while full horizon results, subject-level stability, reliability diagnostics, and statistical analyses are provided in Appendix~\ref{sec:appx_statistical_reliability}.

\subsection{Ablation Studies}
\label{subsec:ablation}

\begin{table}[t]
\centering
\setlength{\tabcolsep}{8.0pt}
\renewcommand{\arraystretch}{1.08}
\small
\caption{
Input-source ablation under $H{=}20$ forecasting.
Each entry reports $p_{\mathrm{corr}}$.
The stimulus-only variant removes brain-history tokens and predicts future brain states from observed stimulus tokens alone.
}
\label{tab:input_source_ablation}
\begin{tabular}{@{}lccc@{}}
\toprule
Variant & Algonauts~2025 & CineBrain & HAD \\
\midrule
Stimulus only            & 0.1391 & 0.3136 & 0.1908 \\
Past fMRI only           & 0.1764 & 0.4872 & 0.2761 \\
Past fMRI + stimulus     & \textbf{0.2090} & \textbf{0.5643} & \textbf{0.2950} \\
\bottomrule
\end{tabular}
\vskip -0.1in
\end{table}

\textbf{Input-source Ablation.}
Table~\ref{tab:input_source_ablation} separates the effects of exogenous stimulus context and endogenous brain-state history.
The stimulus-only variant corresponds to conventional stimulus-to-brain prediction, while the past-fMRI-only variant tests whether forecasting is dominated by temporal continuity.
Combining past fMRI and aligned stimulus context yields the best $p_{\mathrm{corr}}$ on all datasets, supporting our formulation of naturalistic fMRI as an evolving state process driven by both internal brain history and external sensory input.

\begin{figure*}[!t]
  \centering
  \includegraphics[width=0.95\textwidth]{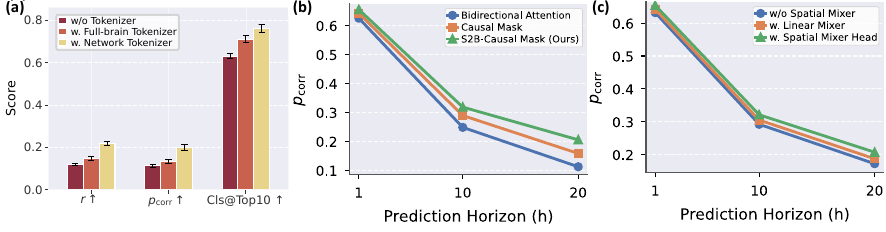}
\caption{\textbf{Ablations on Algonauts~2025.}
(a) Tokenization granularity: continuous fMRI, a shared full-brain tokenizer, and Network-wise Tokenizers.
(b) Temporal-causality masking: bidirectional, standard causal, and S2B-causal attention.
(c) Spatial Mixer Head: no mixer, linear mixer, and the proposed Spatial Mixer Head.}
  \label{fig:ablation}
\vskip -0.12in
\end{figure*}

\textbf{Design Ablations.}
Fig.~\ref{fig:ablation} evaluates the main design choices of BrainVista.
Network-wise tokenization improves over continuous fMRI input and a shared full-brain tokenizer, indicating the value of preserving functional heterogeneity.
S2B improves over vanilla causal masking by $0.008$, $0.023$, $0.040$, $0.056$, and $0.067$ at $H=1,5,10,15,20$, respectively (Table~\ref{tab:s2b_mask_ablation}). Both causal variants restrict future access; S2B additionally blocks brain-to-stimulus attention. The increasing margin is consistent with reduced propagation of prediction errors through stimulus states, but does not directly establish that mechanism.
The Spatial Mixer Head further improves over no-mixer and linear-mixer variants, especially at longer horizons, indicating that explicit cross-network refinement helps reduce recursive error accumulation.
Additional robustness, mechanism, multimodal, and cross-subject analyses are provided in Appendices~\ref{sec:appx_robustness}--\ref{sec:appx_multimodal_transfer}.

\textbf{Joint Mask--Mixer Control.}
Table~\ref{tab:mask_mixer_interaction} jointly varies the temporal mask and attention mixer. At $H=20$, adding the mixer changes Algonauts $p_{\mathrm{corr}}$ by $-0.0153$ under vanilla causal masking and $+0.0278$ under S2B, giving an interaction contrast of $+0.0431$. This sign change shows that the benefit of cross-network refinement depends on the temporal mask in this setting (Appendix~\ref{sec:appx_mask_mixer_interaction}).

\begin{table}[ht]
\centering
\small
\caption{\textbf{Joint mask--mixer control on Algonauts~2025.} Entries are mean $p_{\mathrm{corr}}$ at $H=20$.}
\label{tab:mask_mixer_interaction}
\begin{tabular}{@{}lcc@{}}
\toprule
Temporal mask & No mixer & Attention mixer \\
\midrule
Vanilla causal & 0.1573 & 0.1420 \\
S2B & 0.1812 & \textbf{0.2090} \\
\bottomrule
\end{tabular}
\end{table}

\textbf{Tokenizer Objectives.}
With network partition, latent dimensionality, stimulus encoders, predictor, and rollout protocol fixed, reconstruction-only tokenization reaches $H=20$ $p_{\mathrm{corr}}=0.2090$ on Algonauts, compared with $0.1938$ for VQ-AE, $0.2031$ for $\beta$-VAE, and $0.2044$ for SoftCLIP. SoftCLIP is slightly better at $H=1$ ($0.5613$ versus $0.5600$), but this advantage does not persist at longer horizons. These results support reconstruction-only training among the tested configurations (Appendix~\ref{sec:appx_tokenizer_regularization}).

\textbf{Stimulus Cutoff and Scaling.}
Restricting stimuli to bins strictly before the queried target gives $p_{\mathrm{corr}}=0.2021$, $0.5520$, and $0.2904$ on Algonauts, CineBrain, and HAD (Appendix~\ref{sec:appx_temporal_hrf}); these scores still exceed Transformer-ARX under its default access. Increasing training data from 25\% to 100\% improves $H=20$ from $0.1698$ to $0.2090$. Increasing model size from Base to Large gives only $+0.0028$ while compute rises from 2.1 to 4.9 GPU-hours per subject (Appendix~\ref{sec:appx_scaling}). These results show positive data scaling and diminishing capacity returns, not a general scaling law.

\subsection{Visualization Analysis}
\label{subsec:vis}

\begin{figure*}[!t]
  \centering
  \includegraphics[width=0.9\textwidth]{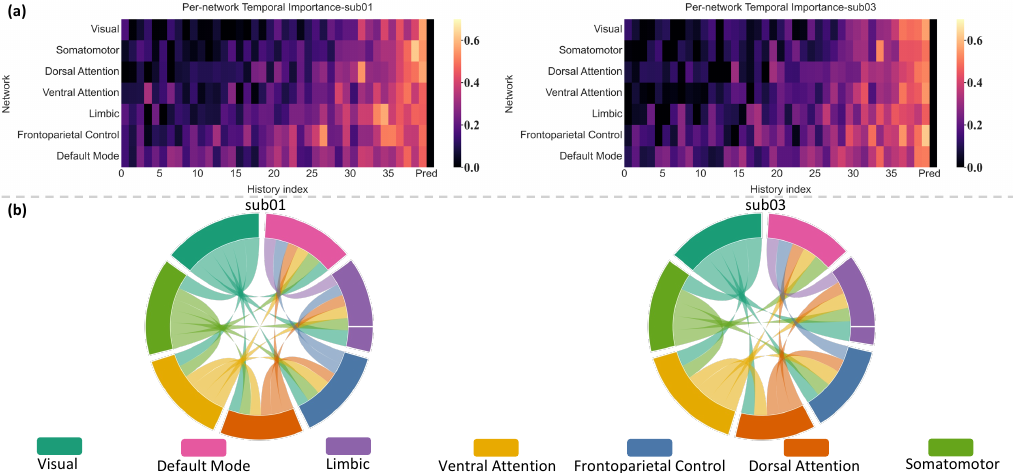}
 \caption{\textbf{Model-based analyses on Algonauts~2025 aggregated to Yeo-7.}
Shown for sub01 and sub03.
(a) History-token occlusion measures the increase in $L_2$ error after masking each history token.
(b) Spatial Mixer attention is aggregated to Yeo-7 network pairs and averaged across time, with chord thickness indicating attention weight.}
  \label{fig:vis_analysis}
\vskip -0.12in
\end{figure*}

\textbf{Temporal Reliance.}
Occluding a history token (Fig.~\ref{fig:vis_analysis}a) while retaining all other history and observed stimuli measures its incremental contribution to held-out prediction. Sensory networks rely more strongly on recent context, whereas association networks retain broader sensitivity to earlier history.
This pattern is consistent with hierarchical intrinsic timescales in cortical processing~\citep{hasson2008hierarchy,lerner2011topographic,murray2014hierarchy,honey2012slow}, and serves as a positive-control validation against established findings. It measures conditional predictive reliance rather than raw autocorrelation, but does not identify a new cortical law or biological mechanism.

\textbf{Cross-network Interaction.}
Spatial Mixer attention exhibits non-uniform coupling across Yeo-7 systems rather than uniform global mixing.
The Spatial Mixer Head thus selectively refines predicted brain tokens through structured network interactions.
These attention patterns describe the fitted model and do not establish biological causality.
Additional visualizations are provided in Appendix~\ref{sec:appx_model_based_analysis}.

\textbf{Auxiliary Representational Utility.}
On held-out Algonauts clips, emotion-category clustering of forecasted brain features improves over BrainHarmony-AR in ARI ($0.138\rightarrow0.214$), NMI ($0.214\rightarrow0.286$), and mean class purity ($0.39\rightarrow0.49$). Ground-truth brain features remain better (Appendix~\ref{sec:appx_downstream_utility}). This tests retention of structure associated with stimulus-derived emotion labels, without assuming near-perfect simulation or validating clinical use.

\section{Conclusion and Discussion}
\label{sec:conclusion}

BrainVista models naturalistic fMRI as observed-stimulus conditional autoregressive forecasting. Its value is a within-run predictive assay for comparing how brain history, stimulus representations, and network organization support held-out recursive prediction. Under matched conditions, Network-wise Tokenizers outperform shared tokenization, S2B improves long-horizon forecasting over vanilla causal masking, and the Spatial Mixer helps under S2B. The positive mask--mixer interaction supports their combined use in the evaluated setting. Forecasted features also retain emotion-category structure in an auxiliary representational test.

Forecasting requires observed initial fMRI and stimuli available through each queried bin, and accuracy decreases with horizon. Cross-subject transfer remains weaker than within-subject prediction. Models are trained separately on each dataset, so the results do not establish zero-shot scanner transfer, arbitrary unseen-movie generalization, or accurate cross-session simulation. History occlusion and attention measure model behavior, not biological causality, connectivity, or a new neural mechanism. The emotion analysis uses stimulus-derived labels and does not validate clinical or individual-level prediction. Future work should test uncertainty calibration, additional data, and broader acquisition settings; further assumptions are discussed in Appendix~\ref{sec:appx_limitations}.

\label{page:main_text_end}
\clearpage
\begin{ack}
This research received no specific external funding. The authors declare no competing interests.
\end{ack}
\bibliographystyle{unsrtnat}
\bibliography{ref}

\clearpage
\appendix

\FloatBarrier
\section{Evaluation Protocol and Reproducibility Details}
\label{sec:appx_protocol}

This appendix specifies the experimental protocol used for all main results, including dataset splits, temporal alignment, conditional autoregressive rollout, training hyperparameters, evaluation metrics, baseline adaptation, dimensionality choices, and computational cost.
We study conditional autoregressive forecasting: after rollout initialization, future ground-truth fMRI states are unavailable, while stimulus tokens are provided according to the same temporally causal interleaved context used by the main model.

\subsection{Dataset Metadata, Splits, and Forecasting Windows}
\label{sec:appx_dataset_metadata}

Table~\ref{tab:dataset_metadata} summarizes the datasets, raw brain-space representations, and model-space representations used by BrainVista.
The three datasets differ in temporal resolution, stimulus paradigm, preprocessing pipeline, and brain-space definition.
For each dataset, we follow the official train/validation/test distribution at the run or episode level.
Validation runs are used only for model selection, and all final numbers are reported on held-out test runs.
Forecasting windows are constructed within individual runs or episodes, and we never carry context tokens, hidden states, predicted states, or normalization statistics across boundaries.

For Algonauts~2025, we use the four challenge subjects: sub-01, sub-02, sub-03, and sub-05.
Each fMRI frame is represented by 1,000 Schaefer parcels, which are mapped to Yeo-7 networks for network-wise tokenization and reporting.
For CineBrain, we use the dataset-defined visual and auditory ROI groups.
Because CineBrain provides high-dimensional voxel-level responses, we reduce each sensory ROI group using PCA fitted only on training runs before tokenization.
For HAD, we use the released surface-based fMRI responses to human-action video clips and aggregate them to Schaefer-1000 parcels before Yeo-7 network grouping.
For start positions with fewer than $L$ previous fMRI steps, the window is discarded rather than padded, ensuring a fixed context length for all evaluated samples.

\subsection{Temporal Alignment, Modality Availability, and Leakage Control}
\label{sec:appx_temporal_alignment}

Stimulus streams are sampled at higher temporal resolution than fMRI and have modality-specific timestamps.
For each dataset $\mathcal{D}$, we define an active modality set:
\begin{equation}
\mathcal{M}_{\mathcal{D}}
\subseteq
\{\mathrm{vis},\mathrm{aud},\mathrm{text}\}.
\end{equation}
Only modalities in $\mathcal{M}_{\mathcal{D}}$ are used to construct stimulus tokens for that dataset.
If a modality is unavailable for an entire dataset, its modality ratio is set to zero and it is excluded from the concatenated stimulus representation.

For each active modality $m\in\mathcal{M}_{\mathcal{D}}$, let $\tau'_{m,i}\in\mathbb{R}^{d_m}$ denote the $i$-th high-rate stimulus token.
For fMRI time step $t$, let:
\begin{equation}
\mathcal{B}_{m,t}
=
\{i:\tau'_{m,i}\text{ is assigned to fMRI bin }t\}
\end{equation}
denote the modality-specific temporal bin.
The fMRI-aligned modality token is computed by mean pooling:
\begin{equation}
\tilde{\tau}_{m}^{t}
=
\frac{1}{|\mathcal{B}_{m,t}|}
\sum_{i\in\mathcal{B}_{m,t}}
\tau'_{m,i},
\qquad
m\in\mathcal{M}_{\mathcal{D}}.
\label{eq:appx_stim_bin_pool}
\end{equation}

The active modality tokens are concatenated in a fixed order and projected into the shared Transformer token space:
\begin{equation}
\tau_s^t
=
\mathrm{Proj}_{\mathcal{M}_{\mathcal{D}}}
\left(
\operatorname*{Concat}_{m\in\mathcal{M}_{\mathcal{D}}}
\tilde{\tau}_m^t
\right)
\in
\mathbb{R}^{d_{\mathrm{model}}},
\label{eq:appx_shared_stim_token}
\end{equation}
where:
\begin{equation}
\mathrm{Proj}_{\mathcal{M}_{\mathcal{D}}}:
\mathbb{R}^{\sum_{m\in\mathcal{M}_{\mathcal{D}}} d_m}
\rightarrow
\mathbb{R}^{d_{\mathrm{model}}}.
\label{eq:appx_active_projection}
\end{equation}
Thus, dataset-level missing modalities are handled by reducing the projection input dimensionality rather than padding unavailable modalities.
For example, when audio is unavailable, the audio ratio is set to zero and the projection operates on the remaining active modality dimensions.

\begin{table}[t]
\centering
\setlength{\tabcolsep}{5.0pt}
\renewcommand{\arraystretch}{1.08}
\small
\caption{\textbf{Stimulus modality availability.}
Dataset-level unavailable modalities are excluded from the concatenation by setting their modality ratio to zero.
The active input dimension is reported after modality-specific feature compression.}
\label{tab:modality_availability}
\begin{tabular}{@{}lccccc@{}}
\toprule
\textbf{Dataset}
& \textbf{Visual}
& \textbf{Audio}
& \textbf{Text}
& \textbf{Active input dim.}
& \textbf{Projection output} \\
\midrule
Algonauts 2025
& Yes
& Yes
& Yes
& $512+256+512=1280$
& $d_{\mathrm{model}}=256$ \\
CineBrain
& Yes
& Yes
& Yes
& $512+256+512=1280$
& $d_{\mathrm{model}}=256$ \\
HAD
& Yes
& No
& Yes
& $512+512=1024$
& $d_{\mathrm{model}}=256$ \\
\bottomrule
\end{tabular}
\vskip -0.1in
\end{table}

For next-step prediction at queried time $u$, the valid conditioning set is:
\begin{equation}
\mathcal{C}_{u}
=
\left(
x^{<u}, s^{\le u}
\right)
=
\left(
(x^1,\ldots,x^{u-1}),
(s^1,\ldots,s^u)
\right).
\label{eq:appx_valid_conditioning_set}
\end{equation}
The target-aligned stimulus token $s^u$ is available because it corresponds to the queried fMRI frame.
However, the target brain state $x^u$, future brain states $x^{>u}$, and future stimulus tokens $s^{>u}$ are never provided as model inputs.
During rollout, this rule is applied recursively: at step $u=T+h$, BrainVista may condition on previously predicted brain states $\hat{x}^{T+1:u-1}$ and observed stimulus tokens $s^{\le u}$, but never on ground-truth future fMRI states.

The Stimulus-to-Brain mask implements this input-access rule in the interleaved sequence.
Stimulus tokens remain in a stimulus-only causal stream, while brain-token queries may attend only to causally available stimulus tokens and available past or predicted brain tokens.
This prevents future fMRI leakage and stimulus look-ahead beyond the queried prediction time.

\subsection{Temporal Validity and HRF Diagnostics}
\label{sec:appx_temporal_hrf}

\begin{table}[t]
\centering
\setlength{\tabcolsep}{1.5pt}
\renewcommand{\arraystretch}{1.10}
\scriptsize
\caption{\textbf{Dataset metadata and model-space representation.}
Raw brain-space dimensionality follows the corresponding dataset release, while model-space dimensionality denotes the representation used by BrainVista after dataset-specific spatial preprocessing.}
\label{tab:dataset_metadata}
\begin{tabular}{@{}lccp{3.2cm}p{3.4cm}p{3.2cm}@{}}
\toprule
\textbf{Dataset} & \textbf{\#Subjects} & \textbf{Native TR} & \textbf{Raw brain representation} & \textbf{Model-space representation} & \textbf{Split and rollout protocol} \\
\midrule
Algonauts 2025
& 4
& 1.49s
& 1,000 Schaefer parcels in MNI space
& 1,000 parcels mapped to Yeo-7 systems; $d_{\mathrm{brain}}=803$
& Official train/validation/test split; rollout within run; context reset at run boundaries. \\

CineBrain
& 6
& 0.8s
& HCP-MMP ROI voxels: 8,405 visual voxels and 10,541 auditory voxels
& Training-set PCA to 512 visual + 512 auditory components; $d_{\mathrm{brain}}=820$
& Official train/validation/test split; audiovisual stimulus and fMRI aligned within run or episode. \\

HAD
& 30
& 2.0s
& Surface-based fMRI responses from participants viewing human-action clips
& fsLR-to-Schaefer-1000 aggregation mapped to Yeo-7 systems; $d_{\mathrm{brain}}=803$
& Official train/validation/test split; rollout within run; context reset at run boundaries. \\
\bottomrule
\end{tabular}
\end{table}

The main forecasting protocol treats the stimulus stream as an observed exogenous input, but all stimulus features must respect the causal cutoff used to construct the interleaved context.
For a queried target time $u$, BrainVista uses brain states strictly before $u$ and stimulus features available by the end of bin $u$. During rollout, earlier predicted states replace unavailable ground truth; target and future ground-truth fMRI are never accessed.
Because BOLD responses are delayed and temporally integrated, we further examine whether different stimulus cutoffs or HRF-aware stimulus constructions affect forecasting performance.
We conduct four diagnostics: timestamp-boundary verification, stimulus lag sweep, HRF/FIR feature construction, and shifted-stimulus negative controls.

\paragraph{Timestamp Convention and Feature Windows.}
For all datasets, we define the stimulus context boundary as the end time of the queried fMRI acquisition bin.
Stimulus features assigned to token $\tau_s^u$ must have feature-window end time no later than this boundary.
Thus, stimulus conditioning uses content observed up to the queried bin boundary, not stimulus content after that boundary.
For datasets whose metadata use center timestamps, we convert them to the same end-time convention by shifting the bin boundary by half a TR.

\begin{table*}[t]
\centering
\setlength{\tabcolsep}{1.5pt}
\renewcommand{\arraystretch}{1.08}
\scriptsize
\caption{\textbf{Timestamp convention and stimulus feature extraction.}
All feature windows are end-aligned: a feature is assigned to $\tau_s^u$ only if its window end time is no later than the queried fMRI bin boundary $t_u$.}
\label{tab:timestamp_feature_windows}
\begin{tabular}{@{}lccp{3.1cm}p{3.1cm}p{3.4cm}@{}}
\toprule
\textbf{Dataset} & \textbf{TR} & \textbf{fMRI boundary} & \textbf{Visual feature window} & \textbf{Audio feature window} & \textbf{Text feature source and boundary} \\
\midrule
Algonauts 2025
& 1.49s
& Bin end
& V-JEPA~2 clips sampled over 2.0s windows; window end $\leq t_u$
& Wav2Vec2-BERT windows of 1.0s with 0.5s stride; window end $\leq t_u$
& Dataset-provided subtitles or aligned transcripts; only text tokens ending before $t_u$ are used. \\

CineBrain
& 0.8s
& Bin end
& V-JEPA~2 clips sampled over 2.0s windows; window end $\leq t_u$
& Wav2Vec2-BERT windows of 1.0s with 0.4s stride; window end $\leq t_u$
& Aligned movie transcript/subtitles; sentence embeddings use causal prefixes only. \\

HAD
& 2.0s
& Bin end
& V-JEPA~2 clips sampled over 2.0s windows; window end $\leq t_u$
& Not used
& Clip-level action descriptions and available annotations are aligned causally to clip boundaries. \\
\bottomrule
\end{tabular}
\end{table*}

\paragraph{Stimulus Lag Sweep.}
We evaluate whether performance depends on the exact stimulus cutoff.
For a cutoff $\Delta$ seconds relative to the queried fMRI bin boundary $t_u$, $\Delta=0$ corresponds to the default causal context setting, $\Delta<0$ restricts the model to earlier stimulus context, and $\Delta>0$ is a diagnostic future-lookahead condition that is not used in the main results.
Table~\ref{tab:stim_lag_sweep} reports $p_{\mathrm{corr}}$ at $H=20$.

\begin{table}[t]
\centering
\setlength{\tabcolsep}{4.3pt}
\renewcommand{\arraystretch}{1.08}
\footnotesize
\caption{\textbf{Stimulus lag sweep at $H=20$.}
Negative cutoffs restrict stimulus access to earlier content; $\Delta=0$ is the default causal context setting; $\Delta=+2$s is a diagnostic lookahead condition and is not used in the main protocol.}
\label{tab:stim_lag_sweep}
\begin{tabular}{@{}lcccccc@{}}
\toprule
\textbf{Dataset} & $\Delta=-8$s & $\Delta=-6$s & $\Delta=-4$s & $\Delta=-2$s & $\Delta=0$s & $\Delta=+2$s \\
\midrule
Algonauts 2025 & 0.185 & 0.193 & 0.199 & 0.205 & \textbf{0.209} & 0.213 \\
CineBrain      & 0.523 & 0.536 & 0.548 & 0.558 & \textbf{0.564} & 0.568 \\
HAD            & 0.275 & 0.281 & 0.288 & 0.292 & \textbf{0.295} & 0.296 \\
\bottomrule
\end{tabular}
\end{table}

The lag sweep shows that performance improves as stimulus context approaches the queried fMRI bin boundary, consistent with the stimulus stream providing useful exogenous information.
However, diagnostic future lookahead yields only marginal additional improvement, suggesting that the main gain is not driven by access to post-boundary stimulus content.

\paragraph{HRF-convolved and FIR Stimulus Features.}
We compare simple bin-wise average pooling with HRF-convolved stimulus features and finite impulse response (FIR) lag features.
For the HRF setting, stimulus features are convolved with a canonical hemodynamic response kernel before being aggregated to the fMRI grid.
For the FIR setting, we concatenate stimulus features from a causal lag bank ending at the queried bin boundary.
No FIR feature uses content after the stimulus context boundary.

\begin{table}[t]
\centering
\setlength{\tabcolsep}{4.8pt}
\renewcommand{\arraystretch}{1.08}
\footnotesize
\caption{\textbf{Stimulus construction ablation at $H=20$.}
All settings are temporally causal with respect to the queried fMRI bin boundary.}
\label{tab:hrf_fir_ablation}
\begin{tabular}{@{}lccc@{}}
\toprule
\textbf{Stimulus construction} & \textbf{Algonauts 2025} & \textbf{CineBrain} & \textbf{HAD} \\
\midrule
Bin-wise average pooling & 0.2090 & 0.5643 & 0.2950 \\
Canonical HRF-convolved features & 0.2075 & 0.5618 & 0.2938 \\
Causal FIR lag bank & \textbf{0.2112} & \textbf{0.5671} & \textbf{0.2965} \\
\bottomrule
\end{tabular}
\end{table}

The HRF-convolved and FIR variants produce comparable performance to the default bin-wise pooling.
This indicates that the main conclusions are not specific to a single alignment heuristic.
We use bin-wise pooling in the main experiments because it is simple, model-agnostic, and avoids introducing additional HRF-shape assumptions that may differ across datasets and cortical systems.

\paragraph{Target-boundary Stimulus Ablation.}
We further test a stricter setting in which the stimulus token at the queried bin boundary is removed.
The model may use only earlier stimulus tokens before the queried bin boundary.
This ablation distinguishes the default causal stimulus context from a stricter history-only stimulus setting.

\begin{table}[t]
\centering
\setlength{\tabcolsep}{4.8pt}
\renewcommand{\arraystretch}{1.08}
\footnotesize
\caption{\textbf{Target-boundary stimulus ablation at $H=20$.}
The earlier-stimulus setting allows only $s^{<u}$, whereas the default causal context setting allows $s^{\leq u}$ for predicting $x^u$. Here $u$ is the queried target index, not the last observed brain-state index.}
\label{tab:no_target_stimulus_ablation}
\begin{tabular}{@{}lccc@{}}
\toprule
\textbf{Stimulus access} & \textbf{Algonauts 2025} & \textbf{CineBrain} & \textbf{HAD} \\
\midrule
Past fMRI only & 0.1764 & 0.4872 & 0.2761 \\
Past fMRI + stimulus $s^{<u}$ & 0.2021 & 0.5520 & 0.2904 \\
Past fMRI + stimulus $s^{\leq u}$ & \textbf{0.2090} & \textbf{0.5643} & \textbf{0.2950} \\
Diagnostic stimulus lookahead $s^{\leq u+1}$ & 0.2131 & 0.5682 & 0.2961 \\
\bottomrule
\end{tabular}
\end{table}

Removing the stimulus token at the queried bin boundary decreases performance but does not eliminate the gain over the past-fMRI-only baseline.
Thus, BrainVista benefits from both endogenous brain history and temporally aligned exogenous stimulus context.
The default setting should be interpreted as conditional autoregressive brain-state forecasting under causal stimulus context, not unconditional future brain simulation.

\paragraph{Shifted-stimulus Negative Controls.}
To test whether the stimulus stream provides meaningful time-aligned information rather than segment identity shortcuts, we shift the stimulus stream within each run while preserving its marginal distribution.
We use circular shifts with offsets of at least 30 seconds and also evaluate cross-run stimulus replacement within the same dataset split.
Both controls preserve stimulus statistics but break temporal correspondence between stimulus and fMRI.

\begin{table}[t]
\centering
\setlength{\tabcolsep}{4.6pt}
\renewcommand{\arraystretch}{1.08}
\footnotesize
\caption{\textbf{Shifted-stimulus negative controls at $H=20$.}
The controls preserve stimulus statistics but disrupt stimulus--fMRI temporal correspondence.}
\label{tab:shifted_stimulus_control}
\begin{tabular}{@{}lccc@{}}
\toprule
\textbf{Setting} & \textbf{Algonauts 2025} & \textbf{CineBrain} & \textbf{HAD} \\
\midrule
Past fMRI only & 0.1764 & 0.4872 & 0.2761 \\
Aligned stimulus & \textbf{0.2090} & \textbf{0.5643} & \textbf{0.2950} \\
Circular-shifted stimulus & 0.1788 & 0.4915 & 0.2774 \\
Cross-run replaced stimulus & 0.1775 & 0.4894 & 0.2769 \\
\bottomrule
\end{tabular}
\end{table}

The shifted-stimulus controls collapse close to the past-fMRI-only baseline.
This supports that the improvement from stimulus conditioning depends on correct temporal alignment rather than retrieval of stimulus identity or segment-level shortcuts.
\subsection{Dataset Heterogeneity and Training-set Normalization}
\label{sec:appx_dataset_heterogeneity}

We do not resample all datasets to a common absolute sampling rate.
Instead, stimulus features are aligned to each dataset's native fMRI grid, and forecasting windows are defined on that grid.
Consequently, one rollout step corresponds to different physical durations across datasets.
For this reason, absolute scores should be interpreted within each dataset; cross-dataset comparisons are used to assess whether relative trends are robust across temporal resolutions and brain-space definitions.

Unless otherwise stated, all experiments use training-set normalization.
For each subject, dataset, network or ROI group $k$, and model-space dimension $j$, normalization statistics are estimated only from training runs:
\begin{equation}
\mu_{k,j}
=
\frac{1}{|\mathcal{I}_{\mathrm{train}}|}
\sum_{(r,t)\in\mathcal{I}_{\mathrm{train}}}
x_{r,j}^{k,t},
\qquad
\sigma_{k,j}
=
\sqrt{
\frac{1}{|\mathcal{I}_{\mathrm{train}}|}
\sum_{(r,t)\in\mathcal{I}_{\mathrm{train}}}
\left(x_{r,j}^{k,t}-\mu_{k,j}\right)^2
}.
\end{equation}
The same fixed statistics are applied to training, validation, and test samples:
\begin{equation}
\bar{x}_{r,j}^{k,t}
=
\frac{x_{r,j}^{k,t}-\mu_{k,j}}
{\sigma_{k,j}+\epsilon}.
\end{equation}
No validation or test fMRI samples are used to estimate normalization statistics.
Run-wise normalization is evaluated only as an offline diagnostic ablation and is not used as the default protocol.

\subsection{Conditional Autoregressive Rollout}
\label{sec:appx_conditional_rollout}

Given an observed context window ending at time $T$, the model receives ground-truth fMRI tokens up to $T$ and stimulus tokens up to the queried prediction step.
It predicts the next fMRI token, feeds the predicted token back into the context, and repeats this process for subsequent steps.
The stimulus stream remains observed, but future ground-truth fMRI states are withheld after initialization.

For an $H$-step rollout, prediction is defined as:
\begin{equation}
\hat{x}^{T+h}
=
f_{\theta}
\left(
[x^{\leq T};\hat{x}^{T+1:T+h-1}],
s^{\leq T+h}
\right),
\qquad
h=1,\ldots,H.
\end{equation}
When $h=1$, the predicted-token history $\hat{x}^{T+1:T+h-1}$ is empty.
For $h>1$, the model recursively consumes its own predicted brain states.
This protocol evaluates long-horizon conditional forecasting under observed naturalistic stimuli without access to future fMRI targets.

\paragraph{Training--evaluation distinction.}
During training, BrainVista uses teacher-forced one-step prediction.
For transition $t\rightarrow t{+}1$, the input contains brain tokens up to $t$ and stimulus tokens up to $t{+}1$; the target brain token at $t{+}1$ is used only as supervision and is never inserted into the input sequence.
During evaluation, after initialization with the observed context, predicted brain tokens are recursively fed back for future predictions.
This distinction keeps optimization stable while measuring error accumulation under autoregressive rollout.

\subsection{Dataset Statistics and Effective Forecasting Windows}
\label{sec:appx_dataset_statistics}

Table~\ref{tab:dataset_statistics_windows} reports the processed data statistics used to construct forecasting anchors.
All anchors are generated independently within each run or episode before aggregation across subjects.
For a run $r$, let:
\begin{equation}
T_r=\mathrm{safeT}(V_r,A_r,W_r,Z_r)
\end{equation}
denote the aligned valid length after truncating the pooled visual, audio, text, and brain-token streams to their common duration.
Let $m=\max(L,L_b)$, where $L$ is the stimulus context length and $L_b$ is the brain-history context length.
Unless otherwise specified, we use $L=40$, $L_b=40$, and rollout horizon $H=20$.

For one-step teacher-forced training and validation, anchors are generated as:
\begin{equation}
\mathcal{A}^{\mathrm{1step}}_r
=
\{t\in\mathbb{Z}: m-1 \le t \le T_r-2\}.
\end{equation}
Each anchor predicts the next brain token $z_{t+1}$ using brain history within $[t-L_b+1,t+1)$ and stimulus context up to the target-aligned stimulus bin.
Specifically, the stimulus window includes $s_{t+1}$ and is indexed as $[t-L+2,t+2)$ under half-open interval notation.
Thus, the number of one-step anchors in run $r$ is:
\begin{equation}
N^{\mathrm{1step}}_r=\max(0,T_r-m).
\end{equation}

For fixed-horizon rollout evaluation, an anchor must additionally contain all $H$ future target brain states within the same run.
The valid rollout-anchor set is:
\begin{equation}
\mathcal{A}^{H}_r
=
\{t\in\mathbb{Z}: m-1 \le t \le T_r-H-1\},
\end{equation}
with count:
\begin{equation}
N^{H}_r=\max(0,T_r-m-H+1).
\end{equation}
Therefore, boundary-crossing windows are not removed after concatenating runs; rather, they are never generated.
The total number of valid anchors is computed as a run-wise sum:
\begin{equation}
N^{\mathrm{1step}}=\sum_r \max(0,T_r-m),
\qquad
N^{H}=\sum_r \max(0,T_r-m-H+1).
\end{equation}

\begin{table}[t]
\centering
\setlength{\tabcolsep}{3.2pt}
\renewcommand{\arraystretch}{1.08}
\scriptsize
\caption{\textbf{Processed dataset statistics and valid run-wise forecasting anchors.}
Frame counts are reported per subject after preprocessing.
One-step anchors are computed as $\sum_r \max(0,T_r-m)$.
Full-horizon rollout anchors are computed as $\sum_r \max(0,T_r-m-H+1)$ with $L=L_b=40$ and $H=20$.
All anchors are generated independently within each run or episode before aggregation across subjects.}
\label{tab:dataset_statistics_windows}
\begin{tabular}{@{}lccccc@{}}
\toprule
\textbf{Dataset}
& \textbf{\#Subj.}
& \textbf{Units/subj.}
& \textbf{Test frames/subj.}
& \textbf{1-step anchors/subj.}
& \textbf{$H{=}20$ anchors/subj.} \\
\midrule
Algonauts 2025
& 4
& 28 / 4 / 4
& $2{,}580\pm33$
& $2{,}420\pm33$
& $2{,}344\pm33$ \\

CineBrain
& 6
& 18 / 3 / 3
& $8{,}100\pm67$
& $7{,}980\pm67$
& $7{,}923\pm67$ \\

HAD
& 30
& 12 / 2 / 2
& $1{,}200\pm0$
& $1{,}120\pm0$
& $1{,}082\pm0$ \\
\bottomrule
\end{tabular}

\vspace{2pt}
\footnotesize
\emph{Note:} Units/subj. denotes train / validation / test units per subject.
Standard deviations are computed across subjects.
\end{table}

Training and validation frame counts follow the same run-wise preprocessing protocol and are used only for model fitting and selection.
The table focuses on test-frame and anchor counts because these determine the effective forecasting evaluation set. The number of valid anchors differs across datasets because of differences in TR, run duration, and split size.
For every dataset, context states, predicted states, and rollout buffers are reset at run or episode boundaries.
Thus, no forecasting anchor contains fMRI context, target states, or stimulus bins from a different run or stimulus segment.

\subsection{Per-subject Run and Window Summary}
\label{sec:appx_subject_window_summary}

Table~\ref{tab:subject_window_summary} summarizes the per-subject unit counts and valid forecasting anchors.
For HAD, the split structure is identical across subjects after preprocessing, so the table reports the common per-subject configuration.
As in Section~\ref{sec:appx_dataset_statistics}, all anchors are counted run-wise rather than after concatenating test runs.

\begin{table}[t]
\centering
\setlength{\tabcolsep}{3.2pt}
\renewcommand{\arraystretch}{1.08}
\footnotesize
\caption{\textbf{Per-subject run/episode counts and valid forecasting anchors.}
One-step anchors are computed as $\sum_r \max(0,T_r-m)$.
Full-horizon rollout anchors are computed as $\sum_r \max(0,T_r-m-H+1)$ with $L=L_b=40$ and $H=20$.
All counts are obtained independently within each run or episode.}
\label{tab:subject_window_summary}
\begin{tabular}{@{}llccccc@{}}
\toprule
\textbf{Dataset}
& \textbf{Subject}
& \textbf{Train units}
& \textbf{Val units}
& \textbf{Test units}
& \textbf{One-step anchors}
& \textbf{$H{=}20$ anchors} \\
\midrule
\multirow{4}{*}{Algonauts 2025}
& sub-01 & 28 & 4 & 4 & 2,397 & 2,321 \\
& sub-02 & 28 & 4 & 4 & 2,459 & 2,383 \\
& sub-03 & 28 & 4 & 4 & 2,435 & 2,359 \\
& sub-05 & 28 & 4 & 4 & 2,389 & 2,313 \\
\midrule
\multirow{6}{*}{CineBrain}
& sub-01 & 18 & 3 & 3 & 7,951 & 7,894 \\
& sub-02 & 18 & 3 & 3 & 8,004 & 7,947 \\
& sub-03 & 18 & 3 & 3 & 7,883 & 7,826 \\
& sub-04 & 18 & 3 & 3 & 8,065 & 8,008 \\
& sub-05 & 18 & 3 & 3 & 8,033 & 7,976 \\
& sub-06 & 18 & 3 & 3 & 7,944 & 7,887 \\
\midrule
HAD
& 30 subjects & 12 & 2 & 2 & 1,120 per subject & 1,082 per subject \\
\bottomrule
\end{tabular}
\end{table}

\subsection{CineBrain PCA Retained Variance}
\label{sec:appx_cinebrain_pca}

CineBrain contains high-dimensional visual and auditory ROI voxel responses.
To keep tokenizer and projection modules comparable across datasets, PCA is fitted separately within each ROI group using training runs only.
The learned PCA basis is then fixed for validation and test runs.
Table~\ref{tab:cinebrain_pca_variance} reports retained variance for the selected dimensionality.

\begin{table}[t]
\centering
\setlength{\tabcolsep}{5.2pt}
\renewcommand{\arraystretch}{1.08}
\footnotesize
\caption{\textbf{CineBrain PCA retained variance.}
PCA is fitted on training runs only and then fixed for validation and test data.}
\label{tab:cinebrain_pca_variance}
\begin{tabular}{@{}lcccc@{}}
\toprule
\textbf{ROI group} & \textbf{Raw dimension $Q_k$} & \textbf{Retained dimension $P_k$} & \textbf{Explained variance} & \textbf{Latent dimension $d_k$} \\
\midrule
Visual ROI   & 8,405  & 512 & $93.6\%\pm1.2\%$ & 410 \\
Auditory ROI & 10,541 & 512 & $92.1\%\pm1.5\%$ & 410 \\
\midrule
Total        & 18,946 & 1,024 & $92.9\%\pm1.3\%$ & 820 \\
\bottomrule
\end{tabular}
\end{table}

We retain 512 components per ROI group as a trade-off between variance preservation and model size.
Increasing the retained dimensionality above 512 produces only marginal forecasting gains while increasing dataset-specific tokenizer and projection parameters.

\begin{table}[t]
\centering
\setlength{\tabcolsep}{4.8pt}
\renewcommand{\arraystretch}{1.08}
\footnotesize
\caption{\textbf{CineBrain PCA dimensionality sensitivity at $H=20$.}
Each row uses the same forecasting architecture after changing the retained PCA dimension per ROI group.}
\label{tab:cinebrain_pca_sensitivity}
\begin{tabular}{@{}lcccc@{}}
\toprule
\textbf{Components per ROI} & \textbf{Model-space dim.} & \textbf{$d_{\mathrm{brain}}$} & \textbf{Explained variance} & $p_{\mathrm{corr}}$ \\
\midrule
256 & 512  & 410  & 87.1\% & 0.5521 \\
512 & 1,024 & 820  & 92.9\% & \textbf{0.5643} \\
768 & 1,536 & 1,230 & 95.2\% & 0.5650 \\
\bottomrule
\end{tabular}
\end{table}

\subsection{Stimulus Feature Extraction Details}
\label{sec:appx_stimulus_feature_details}

Table~\ref{tab:stimulus_feature_details} provides the stimulus feature extraction protocol.
All features are assigned by feature-window end time, and no feature whose window ends after the queried fMRI boundary is used for prediction at that boundary.

\begin{table*}[t]
\centering
\setlength{\tabcolsep}{0.5pt}
\renewcommand{\arraystretch}{1.08}
\scriptsize
\caption{\textbf{Stimulus feature extraction details.}
Visual and audio features are end-aligned to the queried fMRI boundary. Text features use only tokens whose timestamps end before or at the boundary.}
\label{tab:stimulus_feature_details}
\begin{tabular}{@{}lccp{3.0cm}p{3.0cm}p{3.6cm}@{}}
\toprule
\textbf{Dataset} & \textbf{Video rate} & \textbf{Audio rate} & \textbf{Visual encoder} & \textbf{Audio encoder} & \textbf{Text source and context rule} \\
\midrule
Algonauts 2025
& 30 fps
& 16 kHz
& V-JEPA~2; 2.0s clips; 0.5s stride; end-aligned pooling
& Wav2Vec2-BERT; 1.0s windows; 0.5s stride
& Dataset-provided subtitles/transcripts; causal prefix ending at or before $t_u$. \\

CineBrain
& 25 fps
& 16 kHz
& V-JEPA~2; 2.0s clips; 0.4s stride; end-aligned pooling
& Wav2Vec2-BERT; 1.0s windows; 0.4s stride
& Aligned transcript/subtitles; no sentence tokens after $t_u$ are used. \\

HAD
& 30 fps
& Not used
& V-JEPA~2; 2.0s clips; 0.5s stride; end-aligned pooling
& Not used
& Clip-level action descriptions and annotations aligned to clip boundaries. \\
\bottomrule
\end{tabular}
\end{table*}

For text features, we avoid using future sentence context.
If a sentence crosses the queried fMRI boundary, only the prefix whose timestamp does not exceed the boundary is encoded.
If only clip-level annotations are available, the annotation is assigned to the clip interval and becomes visible only after the corresponding clip start, with no access to future clip annotations.

\subsection{Cls@Top10 Candidate Pool Construction}
\label{sec:appx_cls_candidate_pool}

Cls@Top10 evaluates whether a predicted fMRI segment can identify its matched ground-truth segment among distractors.
For each predicted segment, the candidate pool contains one matched ground-truth segment and distractor segments sampled from the same subject and evaluation split.
Distractors are non-overlapping with the query segment and are sampled from different runs or episodes when possible.
The same candidate pools are used for all methods.

\begin{table}[t]
\centering
\setlength{\tabcolsep}{4.4pt}
\renewcommand{\arraystretch}{1.08}
\footnotesize
\caption{\textbf{Cls@Top10 candidate pool construction.}
Chance level is $10/N$, where $N$ is the candidate pool size.}
\label{tab:cls_candidate_pool}
\begin{tabular}{@{}lcccc@{}}
\toprule
\textbf{Dataset} & \textbf{Pool size $N$} & \textbf{Chance Top10} & \textbf{Distractor source} & \textbf{Subject constraint} \\
\midrule
Algonauts 2025 & 200 & 5.0\% & Same split, different run when available & Same subject \\
CineBrain      & 400 & 2.5\% & Same split, different episode when available & Same subject \\
HAD            & 300 & 3.3\% & Same split, different run or clip block & Same subject \\
\bottomrule
\end{tabular}
\end{table}

Using same-subject candidate pools prevents identification from being driven by subject-specific response scale.
Sampling distractors from the same held-out split also reduces distribution mismatch between matched and distractor segments.
Because shifted-stimulus controls in Table~\ref{tab:shifted_stimulus_control} collapse near the past-fMRI-only baseline, the identification results are unlikely to be explained by stimulus identity alone.

\subsection{Random Seed Variance}
\label{sec:appx_seed_variance}

The main forecasting comparisons and Table~\ref{tab:seed_variance} are averaged over five random seeds. The additional comparisons in Appendices~\ref{sec:appx_mask_mixer_interaction}, \ref{sec:appx_tokenizer_regularization}, and~\ref{sec:appx_scaling} report mean scores only; no across-run uncertainty estimate is claimed for those tables.
The seeds affect model initialization, mini-batch order, dropout, and negative candidate sampling for identification.
Table~\ref{tab:seed_variance} reports seed-level mean and standard deviation for the strongest baseline and BrainVista at $H=20$.

\begin{table}[t]
\centering
\setlength{\tabcolsep}{4.4pt}
\renewcommand{\arraystretch}{1.08}
\footnotesize
\caption{\textbf{Random seed variance at $H=20$.}
Each entry reports mean $\pm$ standard deviation of $p_{\mathrm{corr}}$ across five seeds.}
\label{tab:seed_variance}
\begin{tabular}{@{}lccc@{}}
\toprule
\textbf{Dataset} & \textbf{Strongest baseline} & \textbf{Baseline $p_{\mathrm{corr}}$} & \textbf{BrainVista $p_{\mathrm{corr}}$} \\
\midrule
Algonauts 2025 & BrainHarmony-AR & $0.1887\pm0.0034$ & $\mathbf{0.2090\pm0.0031}$ \\
CineBrain      & BrainHarmony-AR & $0.5156\pm0.0048$ & $\mathbf{0.5643\pm0.0042}$ \\
HAD            & BrainHarmony-AR & $0.2769\pm0.0026$ & $\mathbf{0.2950\pm0.0022}$ \\
\bottomrule
\end{tabular}
\end{table}

The seed-level variance is smaller than the improvement of BrainVista over the strongest baseline on all three datasets.
This indicates that the reported gains are not driven by a favorable random initialization or candidate-pool sample.

\subsection{Training and Architecture Hyperparameters}
\label{sec:appx_hyperparameters}

Unless otherwise stated, all variants use context length $L=40$ and rollout horizons $H\in\{1,10,20\}$.
The temporal predictor is a Pre-LN Transformer with hidden size $d_{\mathrm{model}}=256$, $8$ attention heads, and $4$ layers.
We use GELU activations, an FFN expansion ratio of $4\times$, absolute positional embeddings, dropout $0.1$, and gradient clipping at $0.5$.
Models are trained for 40 epochs with AdamW, learning rate $2\times10^{-4}$, weight decay $0.01$, batch size 32, and PyTorch automatic mixed precision.
Weight decay is not applied to LayerNorm parameters or bias terms.

\begin{table}[t]
\centering
\setlength{\tabcolsep}{5.0pt}
\renewcommand{\arraystretch}{1.08}
\small
\caption{\textbf{BrainVista architecture configuration.}
The same temporal forecasting backbone is used across datasets, while brain-token dimensionalities and active stimulus input dimensions are dataset-specific.}
\label{tab:brainvista_architecture}
\begin{tabular}{@{}lc@{}}
\toprule
\textbf{Component} & \textbf{Setting} \\
\midrule
Brain grouping & Yeo-7 or dataset-defined ROI groups \\
Brain tokenizer & Independent two-layer MLP per network \\
Brain detokenizer & Independent two-layer MLP per network \\
Stimulus encoders & Frozen pretrained encoders \\
Stimulus projection output & $d_{\mathrm{model}}=256$ \\
Temporal backbone & Pre-LN causal Transformer \\
Transformer layers & 4 \\
Transformer attention heads & 8 \\
Feed-forward dimension & 1024 \\
Dropout & 0.1 \\
Context length & $L=40$ \\
Brain-history length & $L_b=40$ \\
Default rollout horizon & $H=20$ \\
S2B mask & Directed stimulus-to-brain causal mask \\
Spatial Mixer & One-layer network-wise self-attention \\
Prediction target & Next whole-brain latent token \\
Optimizer & AdamW \\
Learning rate & $2\times10^{-4}$ \\
Weight decay & $10^{-2}$ \\
Batch size & 32 \\
Training epochs & 40 \\
\bottomrule
\end{tabular}
\vskip -0.1in
\end{table}

\paragraph{Network-wise tokenizer parameterization.}
For each functional network $k$, BrainVista uses an independent tokenizer--detokenizer pair:
\begin{equation}
h_{\phi_k}^{k}:\mathbb{R}^{P_k}\rightarrow\mathbb{R}^{d_k},
\qquad
g_{\psi_k}^{k}:\mathbb{R}^{d_k}\rightarrow\mathbb{R}^{P_k}.
\end{equation}
The parameters are network-specific:
\begin{equation}
\Theta_{\mathrm{tok}}
=
\{(\phi_k,\psi_k)\}_{k=1}^{K}.
\end{equation}
No tokenizer or detokenizer parameters are shared across functional networks.
This allows each network to learn a compact representation adapted to its own dimensionality and signal statistics.

\begin{table}[t]
\centering
\setlength{\tabcolsep}{5.0pt}
\renewcommand{\arraystretch}{1.08}
\small
\caption{\textbf{Network-wise tokenizer configuration.}
Each functional network has an independent tokenizer--detokenizer pair.
The latent dimensionalities $d_k$ are dataset- and network-specific and are fixed before forecasting-model training.}
\label{tab:tokenizer_config}
\begin{tabular}{@{}lc@{}}
\toprule
\textbf{Component} & \textbf{Setting} \\
\midrule
Tokenizer input dimension & $P_k$ \\
Tokenizer output dimension & $d_k$ \\
Detokenizer input dimension & $d_k$ \\
Detokenizer output dimension & $P_k$ \\
Tokenizer architecture & Two-layer MLP \\
Detokenizer architecture & Two-layer MLP \\
Parameter sharing across networks & No \\
Training data for normalization & Training split only \\
Tokenizer objective & Network-wise reconstruction loss \\
\bottomrule
\end{tabular}
\vskip -0.1in
\end{table}

The tokenizer--detokenizer pairs are pretrained by minimizing the network-wise reconstruction loss:
\begin{equation}
\mathcal{L}_{\mathrm{tok}}
=
\sum_t
\sum_{k=1}^{K}
\left\|
\bar{x}^{k,t}
-
g_{\psi_k}^{k}
\left(
h_{\phi_k}^{k}(\bar{x}^{k,t})
\right)
\right\|_2^2 .
\label{eq:appx_tokenizer_loss}
\end{equation}
After tokenizer pretraining, the tokenizer parameters are fixed when training the autoregressive forecasting model.
The forecasting model predicts latent brain tokens in the tokenizer space, and the frozen detokenizers map predicted latent tokens back to the normalized fMRI space.

\paragraph{Brain-token Dimensionality.}
For network-wise tokenization, we distinguish raw brain-space dimensionality $Q_k$ from tokenizer input dimensionality $P_k$.
For parcellated inputs, $P_k=Q_k$.
For high-dimensional voxel-level CineBrain inputs, each sensory ROI group is first reduced using PCA fitted on the training split only.
The fitted PCA basis is then fixed and applied to validation and test runs.
We set $d_k=\lceil 0.8P_k\rceil$, and Table~\ref{tab:brain_token_dims} lists the latent dimension of every group.

\begin{table*}[t]
\centering
\setlength{\tabcolsep}{3.0pt}
\renewcommand{\arraystretch}{1.08}
\scriptsize
\caption{\textbf{Network-wise tokenizer dimensionality.}
$Q_k$ denotes the raw parcel, voxel, or surface feature dimension of group $k$.
$P_k$ denotes the tokenizer input dimension after dataset-specific spatial preprocessing.
$d_k=\lceil 0.8P_k\rceil$ is the integer latent dimension, and $d_{\mathrm{brain}}=\sum_k d_k$.
Parcel counts follow the Schaefer-1000 assignment to the Yeo-7 networks.}
\label{tab:brain_token_dims}
\begin{tabular}{@{}llccccc@{}}
\toprule
\textbf{Dataset} & \textbf{Group} & $Q_k$ & \textbf{Preprocessing} & $P_k$ & $d_k$ & \textbf{Projection} \\
\midrule
\multirow{7}{*}{Algonauts 2025}
& Visual              & 162 & Schaefer parcel input & 162 & 130 & \multirow{7}{*}{$803\rightarrow256$} \\
& Somatomotor         & 194 & Schaefer parcel input & 194 & 156 & \\
& Dorsal Attention    & 122 & Schaefer parcel input & 122 & 98  & \\
& Ventral Attention   & 121 & Schaefer parcel input & 121 & 97  & \\
& Limbic              & 60  & Schaefer parcel input & 60  & 48  & \\
& Frontoparietal Ctrl & 129 & Schaefer parcel input & 129 & 104 & \\
& Default Mode        & 212 & Schaefer parcel input & 212 & 170 & \\
\cmidrule(lr){2-7}
& \textbf{Total} & \textbf{1000} & -- & \textbf{1000} & \textbf{803} & \\

\midrule
\multirow{2}{*}{CineBrain}
& Visual ROI   & 8405  & Training-set PCA & 512 & 410 & \multirow{2}{*}{$820\rightarrow256$} \\
& Auditory ROI & 10541 & Training-set PCA & 512 & 410 & \\
\cmidrule(lr){2-7}
& \textbf{Total} & \textbf{18946} & -- & \textbf{1024} & \textbf{820} & \\

\midrule
\multirow{7}{*}{HAD}
& Visual              & 162 & fsLR-to-Schaefer aggregation & 162 & 130 & \multirow{7}{*}{$803\rightarrow256$} \\
& Somatomotor         & 194 & fsLR-to-Schaefer aggregation & 194 & 156 & \\
& Dorsal Attention    & 122 & fsLR-to-Schaefer aggregation & 122 & 98  & \\
& Ventral Attention   & 121 & fsLR-to-Schaefer aggregation & 121 & 97  & \\
& Limbic              & 60  & fsLR-to-Schaefer aggregation & 60  & 48  & \\
& Frontoparietal Ctrl & 129 & fsLR-to-Schaefer aggregation & 129 & 104 & \\
& Default Mode        & 212 & fsLR-to-Schaefer aggregation & 212 & 170 & \\
\cmidrule(lr){2-7}
& \textbf{Total} & \textbf{1000} & -- & \textbf{1000} & \textbf{803} & \\
\bottomrule
\end{tabular}
\end{table*}

\paragraph{CineBrain Voxel Reduction.}
CineBrain provides high-dimensional visual and auditory ROI voxel responses.
Directly tokenizing the raw voxel dimensions would produce substantially larger tokenizer and projection layers than the parcellated datasets.
We therefore fit PCA separately within the visual and auditory ROI groups using only training runs, retaining 512 components per group.
This yields $P_{\mathrm{vis}}=512$, $P_{\mathrm{aud}}=512$, and $d_{\mathrm{brain}}=820$.
All validation and test responses are projected using the training-fitted PCA bases.

\paragraph{Projection and Spatial Mixer Modules.}
The concatenated brain latent $z_f^t\in\mathbb{R}^{d_{\mathrm{brain}}}$ is mapped to the Transformer hidden space by $\mathrm{Proj}^{\mathrm{in}}_b:\mathbb{R}^{d_{\mathrm{brain}}}\rightarrow\mathbb{R}^{256}$.
The temporal Transformer output is mapped back by $\mathrm{Proj}^{\mathrm{out}}_b:\mathbb{R}^{256}\rightarrow\mathbb{R}^{d_{\mathrm{brain}}}$ before Spatial Mixer refinement and detokenization.
The Spatial Mixer operates over the $K$ functional-network components of a predicted whole-brain latent token.
Given a preliminary latent prediction:
\begin{equation}
\bar{z}_f^{u}
=
[
\bar{z}_f^{u,1};
\ldots;
\bar{z}_f^{u,K}
],
\qquad
\bar{z}_f^{u,k}\in\mathbb{R}^{d_k},
\end{equation}
each network component is first mapped to a shared mixer dimension using a network-specific projection:
\begin{equation}
R^{u}
=
\mathrm{Stack}
\left(
W_{\mathrm{in}}^1\bar{z}_f^{u,1},
\ldots,
W_{\mathrm{in}}^K\bar{z}_f^{u,K}
\right)
\in
\mathbb{R}^{K\times d_z}.
\label{eq:appx_spatial_mixer_input}
\end{equation}
The Spatial Mixer first applies multi-head self-attention over the network-token stack:
\begin{equation}
\tilde{R}^{u} = \mathrm{MHA}(R^{u}).
\end{equation}
After self-attention, each mixed network token is mapped back to its original latent dimensionality using another network-specific projection $W_{\mathrm{out}}^{k}$:
\begin{equation}
\hat{z}_{f}^{u,k} = W_{\mathrm{out}}^{k}\tilde{R}^{u}_{k},
\quad
\hat{z}_{f}^{u} = [\hat{z}_{f}^{u,1}; \ldots; \hat{z}_{f}^{u,K}] \in \mathbb{R}^{d_{\mathrm{brain}}}.
\end{equation}
Thus, the attention operation is applied over functional systems, but the input and output mappings remain network-specific.
This prevents the mixer from treating the $K$ network tokens as exchangeable anonymous tokens and preserves the functional-network identity used by the tokenizers.

\begin{table}[t]
\centering
\setlength{\tabcolsep}{5.0pt}
\renewcommand{\arraystretch}{1.08}
\small
\caption{\textbf{Spatial Mixer configuration.}
The Spatial Mixer refines preliminary network-wise brain tokens through self-attention over functional systems.}
\label{tab:spatial_mixer_config}
\begin{tabular}{@{}lc@{}}
\toprule
\textbf{Component} & \textbf{Default setting} \\
\midrule
Input sequence length & $K$ network tokens \\
Shared mixer dimension & $d_z=128$ \\
Attention heads & 4 \\
Mixer layers & 1 \\
Input projection & Network-specific $W_{\mathrm{in}}^k$ \\
Output projection & Network-specific $W_{\mathrm{out}}^k$ \\
Network identity handling & Network-specific projections \\
Mixer scope & Within-step cross-network refinement \\
Attention diagnostic & $K\times K$ network attention matrix \\
\bottomrule
\end{tabular}
\vskip -0.1in
\end{table}

The $K\times K$ attention matrix from the Spatial Mixer is used only as a model-based diagnostic of learned cross-network interactions.
It should not be interpreted as direct biological causality.

\begin{table}[t]
\centering
\setlength{\tabcolsep}{5.0pt}
\renewcommand{\arraystretch}{1.08}
\scriptsize
\caption{\textbf{Compute resources and wall-clock training time.}
Training time includes tokenizer pretraining and temporal predictor training on a single NVIDIA GeForce RTX 5090 GPU.}
\label{tab:compute_training_time}
\begin{tabular}{@{}lccc@{}}
\toprule
\textbf{Dataset} & \textbf{GPU} & \textbf{Training setup} & \textbf{Total training time} \\
\midrule
Algonauts 2025 & RTX 5090 & Single GPU, mixed precision & 12 h 32 min \\
CineBrain      & RTX 5090 & Single GPU, mixed precision & 17 h 18 min \\
HAD            & RTX 5090 & Single GPU, mixed precision & 22 h 46 min \\
\bottomrule
\end{tabular}
\end{table}
\subsection{Metrics}
\label{sec:appx_metrics}

Let $\hat{\mathbf{Y}}\in\mathbb{R}^{T\times D}$ and $\mathbf{Y}\in\mathbb{R}^{T\times D}$ denote predicted and ground-truth fMRI trajectories over $T$ rollout steps and $D$ parcels, voxels, or ROI features.
Unless otherwise stated, metrics are computed per evaluated rollout segment and then averaged over segments and subjects.

\paragraph{Vectorized Pearson Correlation.}
We compute Pearson correlation over the vectorized predicted and ground-truth trajectories:
\begin{equation}
r
=
\mathrm{corr}
\left(
\mathrm{vec}(\hat{\mathbf{Y}}),
\mathrm{vec}(\mathbf{Y})
\right).
\end{equation}
This measures overall trajectory agreement across both time and brain dimensions.

\paragraph{Pattern Correlation.}
Pattern correlation measures the agreement of instantaneous spatial activity patterns:
\begin{equation}
p_{\mathrm{corr}}
=
\frac{1}{T}
\sum_{t=1}^{T}
\mathrm{corr}
\left(
\hat{\mathbf{Y}}_{t,:},
\mathbf{Y}_{t,:}
\right).
\end{equation}
This metric evaluates whether the model preserves the cross-region spatial structure of each predicted brain state.

\paragraph{Cls@Top10.}
For identification, we use segments of length 20 time steps.
For each predicted segment, we construct a candidate pool of ground-truth segments from the same subject but different runs.
Candidates are ranked by correlation similarity to the predicted segment, and Cls@Top10 reports the fraction of queries for which the matched ground-truth segment appears among the top 10.
The same candidate pools are used for all methods.

\paragraph{Additional Metrics.}
We also report mean squared error (MSE), Spearman's rank correlation $\rho$, and two-versus-two identification accuracy (2v2) as complementary diagnostics.
These metrics are not used as training objectives.

\subsection{Existing Assets and Licenses}
\label{sec:appx_existing_assets}

Table~\ref{tab:existing_assets_licenses} summarizes the existing datasets, pretrained encoders, atlases, and preprocessing resources used in this work.
All third-party assets were used only for research evaluation or frozen feature extraction.
We do not redistribute third-party datasets, pretrained checkpoints, or raw stimulus materials with our submission.
All assets were accessed and used according to their original release terms, license statements, or dataset access conditions at the time of evaluation.

\begin{table*}[t]
\centering
\setlength{\tabcolsep}{2.5pt}
\renewcommand{\arraystretch}{1.08}
\scriptsize
\caption{\textbf{Existing assets and licenses.}
We list the primary third-party datasets, pretrained models, atlases, and preprocessing resources used in this work.
Licenses and access terms follow the corresponding official release pages or model cards at the time of evaluation.
We do not redistribute any third-party data or pretrained model weights.}
\label{tab:existing_assets_licenses}
\begin{tabular}{@{}p{0.19\textwidth}p{0.16\textwidth}p{0.28\textwidth}p{0.20\textwidth}p{0.11\textwidth}@{}}
\toprule
\textbf{Asset} & \textbf{Type} & \textbf{Usage in this work} & \textbf{Source / access condition} & \textbf{License / terms} \\
\midrule
Algonauts 2025 / CNeuroMod challenge data~\citep{gifford2025algonauts}
& Dataset
& Naturalistic fMRI forecasting benchmark; used for training, validation, and held-out test evaluation.
& Official Algonauts 2025 challenge release.
& CC0 \\

CineBrain~\citep{gao2025cinebrain}
& Dataset
& Multimodal naturalistic audiovisual fMRI benchmark; visual and auditory ROI responses are reduced with training-set PCA before tokenization.
& Official public dataset release.
& Apache-2.0 \\

HAD~\citep{zhou2023had}
& Dataset
& Human-action fMRI forecasting benchmark; released surface responses are aggregated to Schaefer-1000 parcels for model-space evaluation.
& OpenNeuro dataset release.
& CC0 \\

V-JEPA 2~\citep{assran2025vjepa2}
& Pretrained video encoder
& Frozen visual feature extractor for video stimulus streams.
& Official pretrained checkpoint / repository.
& MIT; selected utility files under Apache-2.0 \\

W2v-BERT 2.0~\citep{seamless2023}
& Pretrained audio encoder
& Frozen audio feature extractor for auditory stimulus streams.
& Official model card / checkpoint release.
& MIT \\

Llama 3.2 3B~\citep{llamateam2024llama3herd,meta2024llama32}
& Pretrained language model
& Frozen text feature extractor for subtitles, transcripts, or aligned textual descriptions.
& Meta model release.
& Llama 3.2 Community License \\

Schaefer-1000 parcellation~\citep{schaefer2018parcellation}
& Brain atlas
& Used to represent cortical responses and to aggregate HAD responses into a common model-space representation.
& Public research atlas files.
& Original atlas release terms \\

Yeo-7 functional networks~\citep{yeo2011organization}
& Brain atlas / network annotation
& Used to assign Schaefer parcels to functional systems for network-wise tokenization and reporting.
& Public research atlas files.
& Original atlas release terms \\

fMRIPrep~\citep{esteban2019fmriprep}
& Preprocessing software reference
& Used by released derivatives or referenced preprocessing pipelines for existing neuroimaging datasets.
& NiPreps / fMRIPrep release.
& Apache-2.0 \\
\bottomrule
\end{tabular}
\end{table*}

\paragraph{Notation Convention.}
Throughout the paper, superscripted time-index ranges such as $x^{\le t}$, $s^{\le t}$, and $\hat{x}^{T+1:T+h-1}$ denote ordered temporal sequences rather than unordered sets.
Specifically,
\[
x^{\le t}=(x^1,\ldots,x^t),
\qquad
s^{\le t}=(s^1,\ldots,s^t),
\qquad
\hat{x}^{T+1:T+h-1}
=
(\hat{x}^{T+1},\ldots,\hat{x}^{T+h-1}).
\]
Set notation is reserved for unordered collections, such as functional-network partitions $\{\mathcal{V}_k\}_{k=1}^{K}$ or index sets.

\FloatBarrier
\section{Baseline Adaptation and Fair Comparison}
\label{sec:appx_baseline_adaptation}

\subsection{Baseline Groups}
\label{sec:appx_baseline_groups}

We compare with two groups of baselines.
The first group contains non-ARX references originally designed for stimulus-to-brain prediction, brain representation learning, or multimodal neural decoding.
These baselines provide reference points but should not be interpreted as direct leaderboard comparisons under their original tasks.
The second group contains ARX baselines explicitly adapted to our conditional autoregressive forecasting protocol.
All ARX baselines receive the same past-fMRI window and the same aligned stimulus features as BrainVista.

\begin{table*}[t]
\centering
\setlength{\tabcolsep}{2.0pt}
\renewcommand{\arraystretch}{1.12}
\scriptsize
\caption{\textbf{Baseline adaptation under the unified forecasting protocol.}
All ARX baselines use the same context length, stimulus features, rollout rule, and evaluation metrics as BrainVista.
For non-forecasting foundation models, we freeze the representation encoder and train a lightweight forecasting head unless otherwise stated.}
\label{tab:baseline_adaptation}
\begin{tabular}{@{}lp{2.8cm}p{3.3cm}p{5.4cm}@{}}
\toprule
\textbf{Model} & \textbf{Original role} & \textbf{Input under our protocol} & \textbf{Adaptation} \\
\midrule
TRIBE
& Stimulus-to-brain encoding
& Aligned video, audio, and text features
& Used as a stimulus-only reference on the same target frames. No future fMRI state is provided. \\

MindSimulator
& Synthetic fMRI / concept-localization modeling
& Stimulus features and dataset-specific ROI/parcellation space
& Used as a non-ARX reference with its prediction branch adapted to each dataset's output space. No target fMRI value is provided as input. \\

BrainLM
& Self-supervised fMRI representation learning
& Past fMRI sequence
& The encoder is frozen, and a lightweight causal forecasting head is trained on top of the learned representation. \\

BrainJEPA
& Predictive brain representation learning
& Past fMRI sequence
& The encoder is frozen, and representation tokens are passed to the same lightweight forecasting interface. \\

Brain Harmony
& Brain foundation representation model
& Past fMRI sequence
& Used as a non-ARX representation reference with a dataset-specific prediction head. \\

Ridge-ARX
& Linear autoregressive baseline
& Flattened past fMRI tokens and aligned stimulus tokens
& Ridge regression predicts the next fMRI state from the concatenated context window and is recursively rolled out with predicted fMRI states. \\

LSTM-ARX
& Recurrent sequence baseline
& Concatenated fMRI--stimulus tokens
& A causal LSTM models temporal dependencies over the context window and recursively consumes predicted brain tokens during rollout. \\

TemporalCNN-ARX
& Causal temporal convolution baseline
& Concatenated fMRI--stimulus tokens
& A temporal convolutional network with causal padding models local temporal history without access to future fMRI states. \\

Transformer-ARX
& Cross-attention forecasting baseline
& Past fMRI tokens and aligned stimulus tokens
& Past fMRI tokens are used as queries, while aligned stimulus tokens are used as keys and values. The fused representation is passed to an MLP forecasting head. \\

BrainHarmony-AR
& Strong adapted foundation baseline
& Brain Harmony representations, past fMRI, and aligned stimulus tokens
& The Brain Harmony encoder is frozen, and a causal AR forecasting head is trained using the same conditional autoregressive protocol as other ARX baselines. \\
\bottomrule
\end{tabular}
\end{table*}

\begin{table}[t]
\centering
\small
\caption{\textbf{Input access under the evaluation protocol.} Stimulus access ends at the queried target bin. Past fMRI means the available history; target and later ground-truth fMRI are withheld. Non-ARX denotes adapted reference methods, not a shared input regime.}
\label{tab:baseline_input_access}
\begin{tabular}{@{}lccc@{}}
\toprule
Model & Past fMRI & Observed stimulus & Target/future fMRI \\
\midrule
TRIBE, MindSimulator & No & Yes & No \\
BrainLM, BrainJEPA, Brain Harmony & Yes & No & No \\
Ridge-, LSTM-, TemporalCNN-ARX & Yes & Yes & No \\
Transformer-ARX, BrainHarmony-AR & Yes & Yes & No \\
BrainVista & Yes & Yes & No \\
\bottomrule
\end{tabular}
\end{table}

\subsection{Scope of Claims}
\label{sec:appx_scope}

Our comparisons concern the conditional forecasting protocol, rather than the original stimulus-only leaderboard settings of the evaluated datasets.
Instead, we study a different problem: forecasting future fMRI states conditioned on both past brain activity and observed naturalistic stimulus streams.
Therefore, all claims about performance are restricted to the conditional autoregressive forecasting protocol defined in Section~\ref{sec:appx_conditional_rollout}.
Non-ARX references are included as references, while ARX baselines provide the main controlled comparison because they receive the same classes of inputs as BrainVista.

\subsection{Non-ARX References versus Matched-access Forecasting Baselines}
\label{sec:appx_native_vs_matched}

Non-ARX stimulus-to-brain and brain-representation methods are included to provide reference points from related modeling paradigms.
However, these methods were not originally designed for recursive conditional forecasting from past fMRI states.
We therefore distinguish them from matched-access ARX baselines.
The matched-access baselines receive the same input sources as BrainVista and are evaluated under the same autoregressive rollout protocol.
In contrast, non-ARX references are evaluated on the corresponding target frames using their adapted prediction heads and should not be interpreted as strict rollout-stability baselines.

\subsection{Full Horizon Results Including Non-ARX References}
\label{sec:appx_full_horizon_native_refs}

Table~\ref{tab:app_had_forecasting_horizons_full} reports the full HAD horizon comparison, including non-ARX stimulus-to-brain and brain-representation references as well as matched-access autoregressive baselines.
Non-ARX references are included for completeness, but they should not be interpreted as strict recursive fMRI-conditioned rollout baselines because they were not originally designed to consume predicted fMRI states during autoregressive evaluation.
The main-text horizon-degradation table therefore focuses on matched-access autoregressive methods.

\begin{table*}[t]
\centering
\setlength{\tabcolsep}{4.2pt}
\renewcommand{\arraystretch}{1.08}
\small
\caption{
\textbf{Full HAD forecasting results across horizons including non-ARX references.}
Each entry reports $r$ and $p_{\mathrm{corr}}$ for $H\in\{1,10,20\}$.
Non-ARX stimulus-to-brain and representation models are reported as reference points, while ARX baselines are matched-access recursive forecasting methods.
}
\label{tab:app_had_forecasting_horizons_full}
\begin{tabular}{@{}lcc|cc|cc@{}}
\toprule
& \multicolumn{2}{c|}{$H{=}1$}
& \multicolumn{2}{c|}{$H{=}10$}
& \multicolumn{2}{c}{$H{=}20$} \\
\cmidrule(lr){2-3}\cmidrule(lr){4-5}\cmidrule(lr){6-7}
\textbf{Model}
& $r$ & $p_{\mathrm{corr}}$
& $r$ & $p_{\mathrm{corr}}$
& $r$ & $p_{\mathrm{corr}}$ \\
\midrule
\multicolumn{7}{c}{\textit{Non-ARX stimulus-to-brain and representation references}} \\
\midrule

TRIBE
& 0.5711 & 0.5419
& 0.3150 & 0.2845
& 0.2685 & 0.1845 \\

MindSimulator
& 0.5782 & 0.5698
& 0.3210 & 0.2960
& 0.2610 & 0.1960 \\

BrainLM
& 0.5695 & 0.5489
& 0.2785 & 0.2250
& 0.2465 & 0.1450 \\

BrainJEPA
& 0.5971 & 0.5723
& 0.2950 & 0.2810
& 0.2515 & 0.2353 \\

Brain Harmony
& 0.5823 & 0.5747
& 0.3085 & 0.3120
& 0.2642 & 0.2688 \\

\midrule
\multicolumn{7}{c}{\textit{Matched-access autoregressive forecasting baselines}} \\
\midrule
Ridge-ARX
& 0.5740 & 0.5530
& 0.2860 & 0.2675
& 0.2630 & 0.2327 \\

LSTM-ARX
& 0.5815 & 0.5740
& 0.3002 & 0.2940
& 0.2642 & 0.2605 \\

TemporalCNN-ARX
& 0.5869 & 0.5831
& 0.3065 & 0.3072
& 0.2654 & 0.2741 \\

Transformer-ARX
& 0.5890 & 0.5876
& 0.3112 & 0.3160
& 0.2678 & 0.2724 \\

BrainHarmony-AR
& 0.5912 & 0.5925
& 0.3136 & 0.3215
& 0.2685 & 0.2769 \\

\midrule
BrainVista
& 0.6058 & 0.6214
& 0.3278 & 0.3390
& 0.2739 & 0.2950 \\
\bottomrule
\end{tabular}
\vskip -0.1in
\end{table*}

\subsection{Baseline Capacity and Adaptation Details}
\label{sec:appx_baseline_capacity}

Transformer-ARX and BrainVista use identical frozen stimulus encoders and pretrained Network-wise Tokenizer--detokenizer pairs. They share accessible inputs, data splits, context length, optimization protocol, tuning budget, and recursive rollout evaluation. Both temporal predictors have four layers and hidden dimension 256. Thus, their comparison controls for these representations and training conditions; the masking and Spatial Mixer ablations separately examine the forecasting computation. Matching the tuning protocol does not imply identical parameter counts or measured compute.

Table~\ref{tab:baseline_capacity} summarizes the capacity and adaptation protocol of the matched-access forecasting baselines.
Parameter counts refer to forecasting-stage trainable parameters after dataset-specific dimensions are instantiated.
Frozen external stimulus encoders, frozen pretrained representation backbones, and frozen tokenizer--detokenizer parameters are excluded from these counts.
A/C/H denote Algonauts~2025, CineBrain, and HAD, respectively.
Parameter counts vary slightly across datasets because the brain-token dimensionality and active stimulus input dimensionality differ.

\begin{table*}[t]
\centering
\setlength{\tabcolsep}{2.2pt}
\renewcommand{\arraystretch}{1.08}
\scriptsize
\caption{\textbf{Baseline capacity and adaptation details.}
All matched-access ARX baselines use past fMRI states and observed stimulus tokens up to the queried prediction step.
Parameter counts are forecasting-stage trainable parameters after dataset-specific dimensions are instantiated and exclude frozen external encoders.}
\label{tab:baseline_capacity}
\begin{tabular}{@{}p{0.135\textwidth}ccccc p{0.18\textwidth} p{0.30\textwidth}@{}}
\toprule
\textbf{Model}
& \textbf{fMRI}
& \textbf{Stim.}
& \textbf{Layers}
& \textbf{Hidden}
& \textbf{Params A/C/H}
& \textbf{Adaptation / protocol} \\
\midrule
Ridge-ARX
& Yes
& Yes
& 1
& --
& $34.02$M / $35.29$M / $34.02$M
& Regularized linear ARX over flattened context windows \\

LSTM-ARX
& Yes
& Yes
& 2
& 256
& $1.67$M / $1.69$M / $1.60$M
& Recurrent model over projected fMRI--stimulus tokens \\

TemporalCNN-ARX
& Yes
& Yes
& 4
& 256
& $1.38$M / $1.40$M / $1.31$M
& Causal temporal convolution over matched input sequence \\

Transformer-ARX
& Yes
& Yes
& 4
& 256
& $3.70$M / $3.72$M / $3.64$M
& Transformer forecasting baseline using matched fMRI--stimulus inputs \\

BrainHarmony-AR
& Yes
& Yes
& 2
& 256
& $2.13$M / $2.15$M / $2.10$M
& Frozen Brain Harmony representation with causal forecasting head \\

\bottomrule
\end{tabular}
\vskip -0.1in
\end{table*}

The Ridge-ARX model has a larger parameter count because it operates on flattened context windows.
For example, with $L=40$, $d_{\mathrm{brain}}=803$, and $d_{\mathrm{model}}=256$, its linear map has $40(803+256)\times803=34.02$M weights.
In contrast, the Transformer-based baselines first project fMRI and stimulus streams into a fixed hidden space and therefore have substantially smaller forecasting-stage parameter counts.

\FloatBarrier
\section{Full Quantitative Results and Statistical Reliability}
\label{sec:appx_statistical_reliability}

\subsection{Full Horizon Breakdown on Algonauts 2025 and CineBrain}
\label{sec:appx_full_horizon}

Table~\ref{tab:appx_rollout_horizons_alg_cine_full_revised} reports the full horizon breakdown on Algonauts 2025 and CineBrain.
The $H=20$ values match the main paper.
The $H=1$ and $H=10$ values are included to show horizon-dependent degradation under recursive prediction.

\begin{table*}[t]
\centering
\setlength{\tabcolsep}{2.9pt}
\renewcommand{\arraystretch}{1.04}
\scriptsize
\caption{\textbf{Rollout performance at different horizons on Algonauts 2025 and CineBrain.}
We report Pearson $r$ and pattern correlation $p_{\mathrm{corr}}$ under conditional autoregressive rollout at $H\in\{1,10,20\}$.}
\label{tab:appx_rollout_horizons_alg_cine_full_revised}
\begin{tabular}{@{}lcc|cc|cc||cc|cc|cc@{}}
\toprule
& \multicolumn{6}{c||}{\textbf{Algonauts 2025}} & \multicolumn{6}{c}{\textbf{CineBrain}} \\
\cmidrule(lr){2-7}\cmidrule(lr){8-13}
& \multicolumn{2}{c|}{$H=1$} & \multicolumn{2}{c|}{$H=10$} & \multicolumn{2}{c||}{$H=20$}
& \multicolumn{2}{c|}{$H=1$} & \multicolumn{2}{c|}{$H=10$} & \multicolumn{2}{c}{$H=20$} \\
\cmidrule(lr){2-3}\cmidrule(lr){4-5}\cmidrule(lr){6-7}
\cmidrule(lr){8-9}\cmidrule(lr){10-11}\cmidrule(lr){12-13}
\textbf{Model}
& $r$ & $p_{\mathrm{corr}}$
& $r$ & $p_{\mathrm{corr}}$
& $r$ & $p_{\mathrm{corr}}$
& $r$ & $p_{\mathrm{corr}}$
& $r$ & $p_{\mathrm{corr}}$
& $r$ & $p_{\mathrm{corr}}$ \\
\midrule
TRIBE
& 0.3695 & 0.3545 & 0.2045 & 0.1962 & 0.1379 & 0.1323
& 0.7647 & 0.4857 & 0.5714 & 0.3630 & 0.4742 & 0.3012 \\

MindSimulator
& 0.3837 & 0.3960 & 0.2124 & 0.2192 & 0.1432 & 0.1478
& 0.7831 & 0.5217 & 0.5852 & 0.3898 & 0.4856 & 0.3235 \\

BrainLM
& 0.3191 & 0.2221 & 0.1767 & 0.1230 & 0.1191 & 0.0829
& 0.7471 & 0.4454 & 0.5583 & 0.3328 & 0.4633 & 0.2762 \\

BrainJEPA
& 0.4000 & 0.2454 & 0.2214 & 0.1359 & 0.1493 & 0.0916
& 0.7968 & 0.3943 & 0.5954 & 0.2946 & 0.4941 & 0.2445 \\

Brain Harmony
& 0.5002 & 0.3928 & 0.2769 & 0.2174 & 0.1867 & 0.1466
& 0.8703 & 0.5618 & 0.6504 & 0.4198 & 0.5397 & 0.3484 \\

\midrule
Ridge-ARX
& 0.4720 & 0.4250 & 0.2470 & 0.2260 & 0.1762 & 0.1584
& 0.8840 & 0.6400 & 0.6600 & 0.4780 & 0.5483 & 0.4016 \\

LSTM-ARX
& 0.5100 & 0.4610 & 0.2680 & 0.2450 & 0.1905 & 0.1718
& 0.8980 & 0.6800 & 0.6790 & 0.5400 & 0.5614 & 0.4529 \\

TemporalCNN-ARX
& 0.5260 & 0.4730 & 0.2770 & 0.2520 & 0.1968 & 0.1756
& 0.9060 & 0.7040 & 0.6920 & 0.5650 & 0.5732 & 0.4724 \\

Transformer-ARX
& 0.5410 & 0.4900 & 0.2860 & 0.2630 & 0.2026 & 0.1819
& 0.9180 & 0.7320 & 0.7130 & 0.5900 & 0.5897 & 0.4978 \\

BrainHarmony-AR
& 0.5520 & 0.5060 & 0.2980 & 0.2740 & 0.2065 & 0.1887
& 0.9260 & 0.7460 & 0.7240 & 0.6090 & 0.5988 & 0.5156 \\

BrainVista
& \textbf{0.5849} & \textbf{0.5600} & \textbf{0.3238} & \textbf{0.3100} & \textbf{0.2183} & \textbf{0.2090}
& \textbf{0.9480} & \textbf{0.9100} & \textbf{0.7610} & \textbf{0.6800} & \textbf{0.6315} & \textbf{0.5643} \\
\bottomrule
\end{tabular}
\end{table*}

\subsection{Full Horizon Breakdown on HAD}
\label{sec:appx_had_full_horizon}

The matched-access HAD results at $H=1,10,20$ are reported in Table~\ref{tab:had_forecasting_horizons}. Table~\ref{tab:app_had_forecasting_horizons_full} additionally includes the non-ARX references under their stated input-access conditions. We use the same values for each method and horizon throughout the paper.

\subsection{Subject-level Stability}
\label{sec:appx_subject_level_stability}
Subject-level mean and standard deviation, paired improvements, and bootstrap intervals are reported together in Table~\ref{tab:subject_level_stability} and Appendix~\ref{sec:appx_subject_level_deltas}. We focus on $p_{\mathrm{corr}}$ at $H=20$ to assess whether long-horizon gains persist across participants.

\subsection{Reliability and Noise-ceiling Diagnostics}
\label{sec:appx_reliability_noise_ceiling}

Absolute fMRI forecasting scores depend on dataset-specific response reliability, preprocessing, spatial resolution, and subject-level signal-to-noise ratio.
We therefore report reliability and noise-ceiling diagnostics to contextualize the magnitude of the predicted response correlations.
These diagnostics are used only for interpretation and are not used for model selection or training.

When repeated or matched held-out response segments are available, we estimate split-half reliability by dividing responses into two disjoint halves, averaging responses within each half, and computing pattern correlation between the two averaged responses.
Let $\rho_{\mathrm{split}}$ denote the raw split-half reliability.
We apply Spearman--Brown correction:
\begin{equation}
\rho_{\mathrm{SB}}
=
\frac{2\rho_{\mathrm{split}}}{1+\rho_{\mathrm{split}}},
\label{eq:spearman_brown}
\end{equation}
where $\rho_{\mathrm{SB}}$ estimates the reliability of the full response set.
We use $\rho_{\mathrm{SB}}$ as a dataset-level reliability ceiling for interpreting $p_{\mathrm{corr}}$.

We further report the ceiling-normalized forecasting score:
\begin{equation}
\mathrm{NC\text{-}Norm}
=
\frac{p_{\mathrm{corr}}(\mathrm{model})}
{\rho_{\mathrm{SB}}},
\label{eq:noise_ceiling_norm}
\end{equation}
where $\mathrm{NC\text{-}Norm}$ measures the fraction of reliable response structure captured by the model.
Because datasets differ in TR, brain-space definition, stimulus protocol, and response reliability, these normalized scores should be interpreted within each dataset rather than as a cross-dataset leaderboard.

\begin{table}[t]
\centering
\setlength{\tabcolsep}{4.2pt}
\renewcommand{\arraystretch}{1.08}
\footnotesize
\caption{\textbf{Reliability and noise-ceiling-normalized forecasting performance.}
Split-half reliability is corrected using the Spearman--Brown formula.
NC-Norm reports the fraction of the estimated reliability ceiling captured by BrainVista.}
\label{tab:noise_ceiling}
\begin{tabular}{@{}lcccc@{}}
\toprule
\textbf{Dataset}
& \textbf{BrainVista $p_{\mathrm{corr}}$}
& \textbf{Split-half reliability}
& \textbf{Reliability ceiling}
& \textbf{NC-Norm} \\
\midrule
Algonauts 2025 & 0.2090 & 0.243 & 0.391 & 0.535 \\
CineBrain      & 0.5643 & 0.586 & 0.739 & 0.764 \\
HAD            & 0.2950 & 0.338 & 0.505 & 0.584 \\
\bottomrule
\end{tabular}
\end{table}

Table~\ref{tab:noise_ceiling} shows that the absolute forecasting scores should be interpreted relative to dataset reliability.
CineBrain has the highest estimated reliability ceiling, consistent with its higher absolute $p_{\mathrm{corr}}$.
Algonauts 2025 and HAD have lower reliability ceilings, which partially explains their lower absolute correlations.
After normalizing by the estimated ceiling, BrainVista captures 53.5\%, 76.4\%, and 58.4\% of the reliable response structure on Algonauts 2025, CineBrain, and HAD, respectively.
This suggests that the method preserves a substantial fraction of predictable brain-state structure even when absolute correlations differ across datasets.

\subsection{Subject-level Stability and Paired Differences}
\label{sec:appx_subject_level_deltas}

We evaluate whether the improvement over the strongest matched-access baseline is consistent across subjects.
All subject-level analyses compare BrainVista with BrainHarmony-AR under $H=20$ forecasting.
BrainHarmony-AR is used as the reference baseline because it is the strongest non-BrainVista matched-access forecasting baseline in the main results.
Table~\ref{tab:subject_level_stability} summarizes subject-level mean performance, paired improvements, and bootstrap confidence intervals.

\begin{table}[t]
\centering
\setlength{\tabcolsep}{4.5pt}
\renewcommand{\arraystretch}{1.08}
\small
\caption{\textbf{Subject-level stability under $H{=}20$ forecasting.}
We compare BrainVista with the strongest matched-access baseline, BrainHarmony-AR.
Each entry reports subject-level $p_{\mathrm{corr}}$ mean $\pm$ standard deviation.
$\Delta$ denotes the paired subject-level improvement.
Bootstrap confidence intervals are computed by resampling subjects with replacement.}
\label{tab:subject_level_stability}
\begin{tabular}{@{}lccccc@{}}
\toprule
\textbf{Dataset}
& $\boldsymbol{n}$
& \textbf{BrainHarmony-AR}
& \textbf{BrainVista}
& $\boldsymbol{\Delta}$
& \textbf{95\% CI of $\boldsymbol{\Delta}$} \\
\midrule
Algonauts 2025
& 4
& $0.1887\pm0.0064$
& $0.2090\pm0.0098$
& $0.0203\pm0.0040$
& $[0.0172,\,0.0234]$ \\
CineBrain
& 6
& $0.5156\pm0.0097$
& $0.5643\pm0.0124$
& $0.0487\pm0.0051$
& $[0.0449,\,0.0523]$ \\
HAD
& 30
& $0.2769\pm0.0181$
& $0.2950\pm0.0209$
& $0.0181\pm0.0028$
& $[0.0171,\,0.0191]$ \\
\bottomrule
\end{tabular}
\vskip -0.1in
\end{table}

Table~\ref{tab:subject_values_alg_cine} reports the subject-level values for Algonauts 2025 and CineBrain.
Table~\ref{tab:subject_values_had} reports the corresponding values for HAD.
These values underlie the win counts, bootstrap intervals, and paired statistical tests.

\begin{table}[t]
\centering
\setlength{\tabcolsep}{5.0pt}
\renewcommand{\arraystretch}{1.08}
\small
\caption{\textbf{Subject-level $p_{\mathrm{corr}}$ under $H{=}20$ forecasting on Algonauts 2025 and CineBrain.}
BH-AR denotes BrainHarmony-AR.
$\Delta$ denotes $p_{\mathrm{corr}}(\mathrm{BrainVista}) - p_{\mathrm{corr}}(\mathrm{BH\mbox{-}AR})$.}
\label{tab:subject_values_alg_cine}
\begin{tabular}{@{}llccc@{}}
\toprule
\textbf{Dataset} & \textbf{Subject} & \textbf{BH-AR} & \textbf{BrainVista} & $\boldsymbol{\Delta}$ \\
\midrule
\multirow{4}{*}{Algonauts 2025}
& sub-01 & 0.1814 & 0.1970 & 0.0156 \\
& sub-02 & 0.1920 & 0.2109 & 0.0189 \\
& sub-03 & 0.1856 & 0.2074 & 0.0218 \\
& sub-05 & 0.1958 & 0.2207 & 0.0249 \\
\midrule
\multirow{6}{*}{CineBrain}
& sub-01 & 0.5054 & 0.5466 & 0.0412 \\
& sub-02 & 0.5180 & 0.5631 & 0.0451 \\
& sub-03 & 0.5226 & 0.5704 & 0.0478 \\
& sub-04 & 0.5068 & 0.5568 & 0.0500 \\
& sub-05 & 0.5302 & 0.5831 & 0.0529 \\
& sub-06 & 0.5106 & 0.5658 & 0.0552 \\
\bottomrule
\end{tabular}
\vskip -0.1in
\end{table}

\begin{table}[t]
\centering
\setlength{\tabcolsep}{4.0pt}
\renewcommand{\arraystretch}{1.04}
\scriptsize
\caption{\textbf{Subject-level $p_{\mathrm{corr}}$ under $H{=}20$ forecasting on HAD.}
BH-AR denotes BrainHarmony-AR.
$\Delta$ denotes $p_{\mathrm{corr}}(\mathrm{BrainVista}) - p_{\mathrm{corr}}(\mathrm{BH\mbox{-}AR})$.}
\label{tab:subject_values_had}
\begin{tabular}{@{}lccc|lccc|lccc@{}}
\toprule
\textbf{Subj.} & \textbf{BH-AR} & \textbf{BrainVista} & $\boldsymbol{\Delta}$
& \textbf{Subj.} & \textbf{BH-AR} & \textbf{BrainVista} & $\boldsymbol{\Delta}$
& \textbf{Subj.} & \textbf{BH-AR} & \textbf{BrainVista} & $\boldsymbol{\Delta}$ \\
\midrule
01 & 0.2526 & 0.2652 & 0.0126
& 11 & 0.2640 & 0.2810 & 0.0170
& 21 & 0.2851 & 0.3048 & 0.0197 \\
02 & 0.2507 & 0.2639 & 0.0133
& 12 & 0.2727 & 0.2900 & 0.0173
& 22 & 0.2901 & 0.3100 & 0.0199 \\
03 & 0.2517 & 0.2657 & 0.0140
& 13 & 0.2747 & 0.2923 & 0.0176
& 23 & 0.2973 & 0.3174 & 0.0202 \\
04 & 0.2479 & 0.2626 & 0.0147
& 14 & 0.2793 & 0.2971 & 0.0178
& 24 & 0.2956 & 0.3160 & 0.0205 \\
05 & 0.2547 & 0.2696 & 0.0149
& 15 & 0.2764 & 0.2944 & 0.0180
& 25 & 0.2939 & 0.3146 & 0.0208 \\
06 & 0.2566 & 0.2719 & 0.0154
& 16 & 0.2771 & 0.2954 & 0.0183
& 26 & 0.2969 & 0.3180 & 0.0212 \\
07 & 0.2594 & 0.2752 & 0.0158
& 17 & 0.2787 & 0.2973 & 0.0186
& 27 & 0.3029 & 0.3245 & 0.0216 \\
08 & 0.2599 & 0.2760 & 0.0162
& 18 & 0.2778 & 0.2966 & 0.0188
& 28 & 0.3027 & 0.3246 & 0.0220 \\
09 & 0.2637 & 0.2802 & 0.0165
& 19 & 0.2875 & 0.3065 & 0.0191
& 29 & 0.3029 & 0.3255 & 0.0226 \\
10 & 0.2645 & 0.2812 & 0.0168
& 20 & 0.2833 & 0.3027 & 0.0194
& 30 & 0.3066 & 0.3299 & 0.0233 \\
\bottomrule
\end{tabular}
\vskip -0.1in
\end{table}

\subsection{Bootstrap Confidence Intervals}
\label{sec:appx_bootstrap_ci}

For each dataset, we compute paired subject-level differences
\[
d_i =
p_{\mathrm{corr}}^{\mathrm{BrainVista}}(i)
-
p_{\mathrm{corr}}^{\mathrm{BH\mbox{-}AR}}(i),
\]
where $i$ indexes subjects and BH-AR denotes BrainHarmony-AR.
We estimate 95\% confidence intervals by resampling subjects with replacement and recomputing the mean paired difference.
All bootstrap intervals are therefore bounded by the empirical range of subject-level paired improvements.
For small-$n$ datasets such as Algonauts 2025, bootstrap intervals have limited resolution, so we report them together with exact paired tests and win counts.

The confidence intervals in Table~\ref{tab:subject_level_stability} remain positive across all three datasets.
The gain is largest on CineBrain and remains consistent on HAD despite larger inter-subject variability.
For Algonauts 2025, BrainVista improves over BrainHarmony-AR for all four subjects, but the result should be interpreted cautiously because the dataset contains only four subjects.

\subsection{Statistical Significance Tests}
\label{sec:appx_significance_tests}

In addition to subject-level standard deviations and bootstrap confidence intervals, we conduct planned one-sided paired subject-level tests at $H=20$.
The planned comparison tests whether BrainVista improves over the strongest matched-access baseline, BrainHarmony-AR.
Because Algonauts 2025 has only four subjects, exact subject-level tests have low resolution.
We therefore report exact sign-test and Wilcoxon signed-rank $p$-values together with win counts.

\begin{table}[t]
\centering
\setlength{\tabcolsep}{5.0pt}
\renewcommand{\arraystretch}{1.08}
\small
\caption{\textbf{One-sided paired subject-level significance tests for $H{=}20$ forecasting.}
Win count reports the number of subjects for which BrainVista outperforms BrainHarmony-AR.
Because Algonauts 2025 contains only four subjects, exact non-parametric tests have limited resolution.}
\label{tab:subject_level_significance}
\begin{tabular}{@{}lcccc@{}}
\toprule
\textbf{Dataset}
& \textbf{Win count}
& \textbf{Mean $\boldsymbol{\Delta}$}
& \textbf{Sign test $\boldsymbol{p}$}
& \textbf{Wilcoxon $\boldsymbol{p}$} \\
\midrule
Algonauts 2025 & $4/4$   & $0.0203$ & $0.0625$ & $0.0625$ \\
CineBrain      & $6/6$   & $0.0487$ & $0.0156$ & $0.0156$ \\
HAD            & $30/30$ & $0.0181$ & $<10^{-6}$ & $<10^{-6}$ \\
\bottomrule
\end{tabular}
\vskip -0.1in
\end{table}

The statistical tests are consistent with the bootstrap intervals in Table~\ref{tab:subject_level_stability}.
BrainVista improves over BrainHarmony-AR for every subject across the three datasets.
For Algonauts 2025, the exact tests remain conservative because of the small subject count.
For CineBrain and HAD, the paired tests provide stronger support for the observed gains.
Together with the positive bootstrap intervals, these results indicate that the improvement is not driven by a single subject or dataset-specific outlier.

\subsection{Complementary Metrics}
\label{sec:appx_complementary_metrics}

Table~\ref{tab:supp_metrics_mse_rho_2v2} reports MSE, Spearman's $\rho$, and 2v2 identification accuracy.
These metrics probe complementary aspects of predictive quality: MSE measures scale-sensitive point-wise fidelity, Spearman's $\rho$ measures monotonic agreement, and 2v2 measures segment-level discriminability.
Non-ARX stimulus-to-brain and representation methods are included as reference points, while ARX baselines provide matched-access forecasting comparisons.

\begin{table*}[t]
\centering
\setlength{\tabcolsep}{3.2pt}
\renewcommand{\arraystretch}{1.05}
\small
\caption{\textbf{Supplementary metrics under conditional autoregressive rollout.}
We additionally report MSE (lower is better), Spearman's $\rho$, and 2v2 identification accuracy.}
\label{tab:supp_metrics_mse_rho_2v2}
\begin{tabular}{lccc|ccc|ccc}
\toprule
& \multicolumn{3}{c|}{\textbf{Algonauts 2025}}
& \multicolumn{3}{c|}{\textbf{CineBrain}}
& \multicolumn{3}{c}{\textbf{HAD}} \\
\textbf{Model}
& \textbf{MSE}$\downarrow$ & $\boldsymbol{\rho}\uparrow$ & \textbf{2v2}$\uparrow$
& \textbf{MSE}$\downarrow$ & $\boldsymbol{\rho}\uparrow$ & \textbf{2v2}$\uparrow$
& \textbf{MSE}$\downarrow$ & $\boldsymbol{\rho}\uparrow$ & \textbf{2v2}$\uparrow$ \\
\midrule
TRIBE
& 1.372 & 0.110 & 0.729
& 0.405 & 0.779 & 0.972
& 1.215 & 0.241 & 0.736 \\
MindSimulator
& 1.243 & 0.115 & 0.731
& 0.382 & 0.783 & 0.974
& 1.154 & 0.249 & 0.739 \\
BrainLM
& 1.486 & 0.072 & 0.695
& 0.428 & 0.751 & 0.968
& 1.284 & 0.198 & 0.711 \\
BrainJEPA
& 1.401 & 0.083 & 0.702
& 0.446 & 0.728 & 0.969
& 1.063 & 0.224 & 0.742 \\
Brain Harmony
& 1.058 & 0.140 & 0.760
& 0.338 & 0.790 & 0.981
& 0.741 & 0.239 & 0.811 \\
\midrule
Ridge-ARX
& 1.102 & 0.150 & 0.768
& 0.316 & 0.812 & 0.976
& 0.732 & 0.232 & 0.818 \\
LSTM-ARX
& 1.032 & 0.164 & 0.778
& 0.298 & 0.834 & 0.979
& 0.704 & 0.239 & 0.834 \\
TemporalCNN-ARX
& 1.008 & 0.170 & 0.784
& 0.286 & 0.846 & 0.980
& 0.681 & 0.244 & 0.842 \\
Transformer-ARX
& 0.976 & 0.178 & 0.795
& 0.270 & 0.858 & 0.982
& 0.666 & 0.247 & 0.850 \\
BrainHarmony-AR
& 0.948 & 0.185 & 0.807
& 0.252 & 0.872 & 0.985
& 0.650 & 0.250 & 0.860 \\
\midrule
\textbf{BrainVista}
& \textbf{0.878} & \textbf{0.230} & \textbf{0.852}
& \textbf{0.221} & \textbf{0.931} & \textbf{0.989}
& \textbf{0.625} & \textbf{0.257} & \textbf{0.872} \\
\bottomrule
\end{tabular}
\vskip -0.1in
\end{table*}

\subsection{Trajectory, Spectral, and Drift Metrics}
\label{sec:appx_traj_spectral_drift}

In addition to instantaneous pattern correlation, we evaluate whether predicted trajectories preserve temporal evolution and distributional structure during rollout.
Let $\hat{\mathbf{Y}}\in\mathbb{R}^{H\times D}$ and $\mathbf{Y}\in\mathbb{R}^{H\times D}$ denote predicted and ground-truth rollout trajectories over horizon $H$ and brain dimension $D$.

\paragraph{Temporal Trajectory Correlation.}
We compute temporal correlation per brain dimension and average over dimensions:
\begin{equation}
r_{\mathrm{traj}}
=
\frac{1}{D}
\sum_{j=1}^{D}
\mathrm{corr}
\left(
\hat{\mathbf{Y}}_{:,j},
\mathbf{Y}_{:,j}
\right).
\end{equation}
This metric evaluates whether the model preserves the temporal evolution of each parcel, voxel, or ROI feature.

\paragraph{Distribution Drift.}
We measure rollout drift by comparing predicted and ground-truth first- and second-order statistics:
\begin{equation}
\Delta_{\mu}
=
\frac{1}{H}
\sum_{h=1}^{H}
\left|
\mu(\hat{\mathbf{Y}}_{h,:})-\mu(\mathbf{Y}_{h,:})
\right|,
\qquad
\Delta_{\sigma}
=
\frac{1}{H}
\sum_{h=1}^{H}
\left|
\sigma(\hat{\mathbf{Y}}_{h,:})-\sigma(\mathbf{Y}_{h,:})
\right|.
\end{equation}
Lower values indicate less distributional drift during autoregressive rollout.

\paragraph{Spectral Fidelity.}
Because BOLD responses are temporally smooth, we further compare normalized power spectra between predicted and ground-truth trajectories.
For each evaluated run and network, we compute the log power spectrum of predicted and ground-truth time series and report the mean absolute log-spectral distance:
\begin{equation}
D_{\mathrm{PSD}}
=
\frac{1}{K}
\sum_{k=1}^{K}
\left\|
\log P_k(\hat{\mathbf{Y}})
-
\log P_k(\mathbf{Y})
\right\|_1 .
\end{equation}
Spectral metrics are computed within valid run boundaries and are never computed across concatenated independent runs.

\begin{table}[t]
\centering
\setlength{\tabcolsep}{4.0pt}
\renewcommand{\arraystretch}{1.08}
\footnotesize
\caption{\textbf{Trajectory, spectral, and drift metrics under $H{=}20$ rollout.}
Higher is better for $r_{\mathrm{traj}}$.
Lower is better for mean drift, std drift, and PSD distance.
BH-AR denotes BrainHarmony-AR.}
\label{tab:traj_spectral_drift}
\begin{tabular}{@{}llcccc@{}}
\toprule
\textbf{Dataset}
& \textbf{Model}
& $r_{\mathrm{traj}}\uparrow$
& $\Delta_{\mu}\downarrow$
& $\Delta_{\sigma}\downarrow$
& $D_{\mathrm{PSD}}\downarrow$ \\
\midrule
\multirow{2}{*}{Algonauts 2025}
& BH-AR      & 0.204 & 0.083 & 0.071 & 0.192 \\
& BrainVista & \textbf{0.247} & \textbf{0.067} & \textbf{0.058} & \textbf{0.163} \\
\midrule
\multirow{2}{*}{CineBrain}
& BH-AR      & 0.598 & 0.061 & 0.052 & 0.158 \\
& BrainVista & \textbf{0.641} & \textbf{0.048} & \textbf{0.043} & \textbf{0.132} \\
\midrule
\multirow{2}{*}{HAD}
& BH-AR      & 0.279 & 0.074 & 0.066 & 0.177 \\
& BrainVista & \textbf{0.298} & \textbf{0.063} & \textbf{0.056} & \textbf{0.151} \\
\bottomrule
\end{tabular}
\vskip -0.1in
\end{table}

Table~\ref{tab:traj_spectral_drift} shows that BrainVista improves not only instantaneous spatial-pattern agreement but also trajectory-level temporal fidelity.
The reductions in mean drift, variance drift, and PSD distance indicate that BrainVista better preserves rollout-scale distributional and spectral structure.
These diagnostics complement the main $p_{\mathrm{corr}}$ results and are not used as training objectives.

\FloatBarrier
\section{Robustness to Context Length, Horizon, and Normalization}
\label{sec:appx_robustness}

\begin{table}[t]
\centering
\setlength{\tabcolsep}{6.0pt}
\renewcommand{\arraystretch}{1.08}
\footnotesize
\caption{\textbf{Context-length sensitivity on HAD.}
Each entry reports $p_{\mathrm{corr}}$ under conditional autoregressive rollout.}
\label{tab:context_length_sensitivity}
\begin{tabular}{@{}cccc@{}}
\toprule
\textbf{Context length $L$} & $H=1$ & $H=10$ & $H=20$ \\
\midrule
10 & 0.585 & 0.287 & 0.213 \\
20 & 0.607 & 0.326 & 0.269 \\
40 & 0.621 & 0.339 & 0.295 \\
60 & 0.622 & 0.340 & 0.294 \\
\bottomrule
\end{tabular}
\end{table}

\subsection{Context-length Sensitivity}
\label{sec:appx_context_length}

Table~\ref{tab:context_length_sensitivity} evaluates the effect of context length $L$ on HAD.
Very short context windows reduce long-horizon forecasting performance.
Increasing $L$ from 10 to 40 improves all horizons, while performance saturates when increasing $L$ from 40 to 60.
We therefore use $L=40$ as the default trade-off between temporal coverage and computational cost.

\begin{table}[t]
\centering
\setlength{\tabcolsep}{4.6pt}
\renewcommand{\arraystretch}{1.08}
\footnotesize
\caption{\textbf{Tokenizer reconstruction diagnostics across datasets.}
Metrics are computed in normalized model space on validation runs.
All tokenizer--detokenizer pairs are trained only on training runs.}
\label{tab:tokenizer_reconstruction_diagnostics}
\begin{tabular}{@{}lccc@{}}
\toprule
\textbf{Dataset}
& \textbf{Recon. MSE}$\downarrow$
& \textbf{Recon. $r$}$\uparrow$
& \textbf{Explained variance}$\uparrow$ \\
\midrule
Algonauts 2025 & 0.041 & 0.958 & 0.914 \\
CineBrain      & 0.047 & 0.951 & 0.901 \\
HAD            & 0.044 & 0.955 & 0.908 \\
\bottomrule
\end{tabular}
\vskip -0.1in
\end{table}

\subsection{Normalization Strategy}
\label{sec:appx_normalization}

Table~\ref{tab:normalization_strategy} compares normalization strategies on HAD.
The default setting is training-set per-network normalization, where statistics are computed only from training runs and reused for validation and test rollout.
Subject-level and global normalization are slightly worse, while run-wise per-network normalization is included only as an offline diagnostic.

\begin{table}[t]
\centering
\setlength{\tabcolsep}{5.2pt}
\renewcommand{\arraystretch}{1.08}
\footnotesize
\caption{\textbf{Normalization strategy comparison on HAD.}
Each entry reports $p_{\mathrm{corr}}$. The default setting is training-set per-network normalization.}
\label{tab:normalization_strategy}
\begin{tabular}{@{}lccc@{}}
\toprule
\textbf{Normalization} & $H=1$ & $H=10$ & $H=20$ \\
\midrule
Training-set per-network & \textbf{0.621} & \textbf{0.339} & \textbf{0.295} \\
Subject-level            & 0.614 & 0.331 & 0.287 \\
Global                   & 0.606 & 0.321 & 0.279 \\
Run-wise per-network diagnostic & 0.618 & 0.336 & 0.292 \\
\bottomrule
\end{tabular}
\end{table}
\subsection{Sequence-boundary Handling}
\label{sec:appx_sequence_boundary}

We reset temporal context at run or episode boundaries and never carry history tokens across independent stimulus segments.
This is important because naturalistic datasets often contain multiple stimulus segments, episodes, or runs, and carrying context across boundaries could introduce artificial temporal continuity and inflate forecasting performance.

All forecasting anchors are generated independently within each run or episode.
For a run $r$ with aligned valid length $T_r=\mathrm{safeT}(V_r,A_r,W_r,Z_r)$, let $m=\max(L,L_b)$.
One-step anchors satisfy:
\begin{equation}
m-1 \le t \le T_r-2,
\end{equation}
and fixed-horizon rollout anchors with horizon $H$ additionally satisfy:
\begin{equation}
m-1 \le t \le T_r-H-1.
\end{equation}
Equivalently, the run-wise anchor counts are:
\begin{equation}
N^{\mathrm{1step}}=\sum_r \max(0,T_r-m),
\qquad
N^{H}=\sum_r \max(0,T_r-m-H+1).
\end{equation}
These counts are computed within runs before aggregation across subjects.
Thus, boundary-crossing windows are never generated, rather than being filtered after concatenating runs.

\FloatBarrier
\section{Design and Mechanism Ablations}
\label{sec:appx_design_ablations}

\begin{table}[t]
\centering
\setlength{\tabcolsep}{4.2pt}
\renewcommand{\arraystretch}{1.10}
\footnotesize
\caption{\textbf{Mapping from modeling challenges to BrainVista components.}
Each component is tied to a specific failure mode in naturalistic fMRI forecasting and evaluated by a corresponding diagnostic experiment.}
\label{tab:challenge_component_mapping}
\begin{tabular}{@{}p{3.0cm}p{3.0cm}p{3.8cm}@{}}
\toprule
\textbf{Challenge} & \textbf{BrainVista component} & \textbf{Diagnostic evidence} \\
\midrule
Functional heterogeneity across cortical systems
& Network-wise Tokenizers
& Tokenization granularity ablation and network-level rollout fidelity. \\

Temporal leakage under mismatched stimulus--BOLD timescales
& Stimulus-to-Brain mask
& Mask ablation comparing bidirectional, vanilla causal, and S2B-causal attention. \\

Error accumulation in recursive forecasting
& Conditional autoregressive rollout with past brain-state feedback
& Horizon breakdown and context-length sensitivity. \\

Cross-network interaction without over-smoothing
& Spatial Mixer Head
& Spatial Mixer ablation and model-based attention diagnostics. \\

Dependence on multimodal sensory drive
& Time-aligned multimodal stimulus tokens
& Input-source and modality ablations. \\
\bottomrule
\end{tabular}
\end{table}

\subsection{Tokenization Granularity}
\label{sec:appx_tokenization_granularity}

Table~\ref{tab:tokenization_granularity} compares different brain-state representations on Algonauts 2025 at $H=20$.
Network-wise tokenization improves over both continuous fMRI input and a shared full-brain tokenizer, supporting the need to preserve functional heterogeneity before temporal forecasting.

\begin{table}[t]
\centering
\setlength{\tabcolsep}{5.0pt}
\renewcommand{\arraystretch}{1.08}
\footnotesize
\caption{\textbf{Tokenization granularity ablation on Algonauts 2025 at $H=20$.}}
\label{tab:tokenization_granularity}
\begin{tabular}{@{}lccc@{}}
\toprule
\textbf{Variant} & $r$ & $p_{\mathrm{corr}}$ & Cls@Top10 \\
\midrule
Continuous fMRI input & 0.151 & 0.118 & 0.673 \\
Shared full-brain tokenizer & 0.184 & 0.166 & 0.721 \\
Network-wise tokenizer & \textbf{0.218} & \textbf{0.209} & \textbf{0.762} \\
\bottomrule
\end{tabular}
\end{table}

\subsection{Full Definition of the Stimulus-to-Brain Mask}
\label{sec:appx_s2b_mask}

For queried target time $u$, let $S_u$ be the ordered stimulus--brain sequence in Eq.~\ref{eq:interleave}. It contains the stimulus window through $u$, the brain-history window strictly before $u$, and a target-query position that contains no ground-truth fMRI. The exact context index ranges are specified in Appendix~\ref{sec:appx_dataset_statistics}.
Each token has a time index $t_i$ and a type $a_i\in\{s,b\}$, where $s$ and $b$ denote stimulus tokens and brain tokens, respectively.
The binary Stimulus-to-Brain mask $M_u$ is defined as:
\begin{equation}
M_{ij}=1
\Longleftrightarrow
\begin{cases}
a_i=s,\ a_j=s,\ t_j\leq t_i, \\
a_i=b,\ a_j=s,\ t_j\leq t_i, \\
a_i=b,\ a_j=b,\ t_j\leq t_i .
\end{cases}
\label{eq:appx_s2b_mask}
\end{equation}
All other entries are set to zero.
This mask is lower-triangular in time and enforces directed mixing between the stimulus and brain streams: stimulus tokens do not attend to brain tokens, while brain tokens may attend to both stimulus and brain tokens within the allowed past.
For a queried target $u$, the input contains observed stimulus tokens through $u$ and brain tokens strictly before $u$; the final brain position is a prediction query, not an encoding of $x^u$. The diagonal in Eq.~\ref{eq:appx_s2b_mask} therefore does not grant access to target ground-truth fMRI.

\subsection{Stimulus-to-Brain Mask Ablation}
\label{sec:appx_s2b_mask_ablation}

Table~\ref{tab:s2b_mask_ablation} compares bidirectional attention, a standard mask, and the proposed S2B mask on Algonauts 2025.
Bidirectional attention is not temporally valid for forecasting because it can expose the model to future-aligned context.
A standard mask is better, but it follows only the token order of the interleaved sequence and does not explicitly encode the directed stimulus-to-brain mixing used in BrainVista.
In contrast, the S2B mask prevents stimulus tokens from attending to brain tokens, while allowing brain tokens to integrate both stimulus and brain history within the allowed causal context.
S2B has the highest $p_{\mathrm{corr}}$ at all five evaluated horizons, including the intermediate horizons $H=5$ and $H=15$.
Its absolute advantage over vanilla causal masking increases from $0.008$ at $H=1$ to $0.067$ at $H=20$; the corresponding relative improvements are $1.4\%$ and $47.2\%$.

\begin{table}[t]
\centering
\setlength{\tabcolsep}{5.5pt}
\renewcommand{\arraystretch}{1.08}
\footnotesize
\caption{\textbf{Temporal mask ablation on Algonauts 2025.}
Scores report $p_{\mathrm{corr}}$ under conditional autoregressive rollout; the last row gives the relative improvement of S2B over vanilla causal masking.}
\label{tab:s2b_mask_ablation}
\begin{tabular}{@{}lccccc@{}}
\toprule
\textbf{Mask} & $H=1$ & $H=5$ & $H=10$ & $H=15$ & $H=20$ \\
\midrule
Bidirectional attention & 0.545 & 0.361 & 0.245 & 0.169 & 0.113 \\
Vanilla causal mask     & 0.552 & 0.380 & 0.270 & 0.195 & 0.142 \\
S2B mask                & \textbf{0.560} & \textbf{0.403} & \textbf{0.310} & \textbf{0.251} & \textbf{0.209} \\
\midrule
S2B relative gain       & $1.4\%$ & $6.1\%$ & $14.8\%$ & $28.7\%$ & $47.2\%$ \\
\bottomrule
\end{tabular}
\end{table}

\subsection{Spatial Mixer Head}
\label{sec:appx_spatial_mixer_ablation}

Table~\ref{tab:spatial_mixer_ablation} ablates the Spatial Mixer Head on Algonauts 2025.
The gain is modest for one-step prediction but becomes clearer at longer horizons, suggesting that explicit cross-network interaction modeling helps reduce drift during recursive forecasting.

\begin{table}[t]
\centering
\setlength{\tabcolsep}{5.5pt}
\renewcommand{\arraystretch}{1.08}
\footnotesize
\caption{\textbf{Spatial Mixer Head ablation on Algonauts 2025.}
Each entry reports $p_{\mathrm{corr}}$.}
\label{tab:spatial_mixer_ablation}
\begin{tabular}{@{}lccc@{}}
\toprule
\textbf{Variant} & $H=1$ & $H=10$ & $H=20$ \\
\midrule
w/o Spatial Mixer & 0.548 & 0.294 & 0.181 \\
Linear Mixer      & 0.553 & 0.302 & 0.195 \\
Spatial Mixer Head & \textbf{0.560} & \textbf{0.310} & \textbf{0.209} \\
\bottomrule
\end{tabular}
\end{table}

\subsection{Tokenizer Pretraining vs. End-to-end Training}
\label{sec:appx_pretraining_e2e}

Table~\ref{tab:tokenizer_pretraining_e2e} compares decoupled tokenizer pretraining with end-to-end training on HAD.
End-to-end training is possible but less stable under limited naturalistic fMRI data and shows worse long-horizon performance.
We therefore pretrain the tokenizer separately and freeze it during forecasting training.

\begin{table}[t]
\centering
\setlength{\tabcolsep}{5.5pt}
\renewcommand{\arraystretch}{1.08}
\footnotesize
\caption{\textbf{Tokenizer pretraining vs. end-to-end training on HAD.}
Each entry reports $p_{\mathrm{corr}}$.}
\label{tab:tokenizer_pretraining_e2e}
\begin{tabular}{@{}lccc@{}}
\toprule
\textbf{Training strategy} & $H=1$ & $H=10$ & $H=20$ \\
\midrule
End-to-end training & 0.574 & 0.246 & 0.173 \\
Pretrained tokenizer + frozen encoder & \textbf{0.621} & \textbf{0.339} & \textbf{0.295} \\
\bottomrule
\end{tabular}
\end{table}

\subsection{Tokenizer Reconstruction Diagnostics}
\label{sec:appx_tokenizer_reconstruction}

To isolate tokenizer capacity from forecasting performance, we evaluate standalone reconstruction before training BrainVista.
These results should not be interpreted as final forecasting scores, because forecasting also depends on causal temporal prediction and multimodal conditioning.

\begin{table}[t]
\centering
\setlength{\tabcolsep}{5.2pt}
\renewcommand{\arraystretch}{1.08}
\footnotesize
\caption{\textbf{Tokenizer reconstruction diagnostics.}
Reconstruction correlation $r$ is reported. This is a standalone tokenizer diagnostic, not end-to-end forecasting performance.}
\label{tab:tokenizer_recon_ablation}
\begin{tabular}{@{}llc@{}}
\toprule
\textbf{Family} & \textbf{Objective} & $r$ \\
\midrule
MLP      & AE  & \textbf{0.834} \\
MLP      & VAE & 0.507 \\
BrainVAE & AE  & 0.814 \\
BrainVAE & VAE & 0.797 \\
\bottomrule
\end{tabular}
\end{table}

\subsection{Interaction between Temporal Masking and Spatial Mixing}
\label{sec:appx_mask_mixer_interaction}

We jointly vary the temporal mask and the presence of the attention-based Spatial Mixer on Algonauts~2025 at $H=20$. Table~\ref{tab:mask_mixer_interaction} reports the four controlled configurations. The attention mixer decreases $p_{\mathrm{corr}}$ by $0.0153$ under vanilla causal masking but increases it by $0.0278$ under S2B. The corresponding difference-in-differences is
\[
(0.2090-0.1812)-(0.1420-0.1573)=0.0431.
\]
Thus, the benefit of mixing depends on the temporal information-flow constraint in this experiment. These measurements support the combined design in the evaluated setting; they do not establish a universal advantage of either operator or directly measure the proposed error-propagation pathway.

\subsection{Tokenizer Objective and Latent Regularization}
\label{sec:appx_tokenizer_regularization}

The default tokenizer is a deterministic network-specific autoencoder trained with reconstruction loss and frozen before forecasting. It has no additional distributional prior: the bottleneck, network partition, reconstruction objective, and fixed tokenizer--detokenizer mapping constrain the representation. Absence of a latent prior does not mean that the tokens are random projections.

We compare this reconstruction-only objective with three additional constraints on Algonauts~2025. VQ-AE adds codebook and commitment losses and uses discrete latent codes. The $\beta$-VAE adds a KL penalty toward an isotropic Gaussian prior. AE + SoftCLIP adds alignment to frozen stimulus embeddings. The network partition, latent dimensionality, stimulus encoders, forecasting model, and rollout protocol are held fixed. Table~\ref{tab:tokenizer_regularization} reports reconstruction and forecasting results for these evaluated configurations.

\begin{table}[ht]
\centering
\small
\setlength{\tabcolsep}{5pt}
\caption{\textbf{Tokenizer regularization on Algonauts~2025.} Reconstruction correlation and MSE measure signal fidelity; forecasting columns report mean $p_{\mathrm{corr}}$. Higher correlation and lower MSE are better.}
\label{tab:tokenizer_regularization}
\begin{tabular}{@{}lccccc@{}}
\toprule
Tokenizer & Recon. $r$ & Recon. MSE & $H=1$ & $H=10$ & $H=20$ \\
\midrule
VQ-AE & 0.916 & 0.080 & 0.5384 & 0.2887 & 0.1938 \\
$\beta$-VAE & 0.934 & 0.064 & 0.5521 & 0.3021 & 0.2031 \\
AE + SoftCLIP & 0.947 & 0.052 & \textbf{0.5613} & 0.3062 & 0.2044 \\
Reconstruction only & \textbf{0.958} & \textbf{0.041} & 0.5600 & \textbf{0.3100} & \textbf{0.2090} \\
\bottomrule
\end{tabular}
\end{table}

Reconstruction-only training has the highest reconstruction fidelity and the highest mean forecasting score at $H=10$ and $H=20$. SoftCLIP has a slightly higher mean at $H=1$ ($+0.0013$), which does not persist at longer horizons. These comparisons support the default objective among the tested configurations; they do not imply that latent regularization is generally ineffective or that other regularization settings cannot improve forecasting.

\FloatBarrier
\section{Multimodal Fusion and Cross-subject Transfer}
\label{sec:appx_multimodal_transfer}

\subsection{Modality Ablation}
\label{sec:appx_modality_ablation}

Table~\ref{tab:alg_modality_ablation_pcorr} evaluates the contribution of each stimulus modality on Algonauts~2025.
The full multimodal setting performs best across horizons, indicating that visual, auditory, and textual streams provide complementary information for conditional brain-state forecasting.
Removing video causes the largest degradation, suggesting that visual context contributes strongly to preserving cross-ROI spatial patterns under long-horizon rollout.
Single-modality variants further show that multimodal conditioning is more stable than relying on any individual input stream.

\begin{table}[t]
\centering
\setlength{\tabcolsep}{4.2pt}
\renewcommand{\arraystretch}{1.04}
\footnotesize
\caption{\textbf{Modality ablation on Algonauts~2025.}
Each entry reports $p_{\mathrm{corr}}$.
The full multimodal setting uses visual, audio, and text features.}
\label{tab:alg_modality_ablation_pcorr}
\begin{tabular}{@{}lccc@{}}
\toprule
\textbf{Variant} & $H=1$ & $H=10$ & $H=20$ \\
\midrule
All (V+A+T)     & \textbf{0.560} & \textbf{0.310} & \textbf{0.209} \\
w/o Video (A+T) & 0.510 & 0.270 & 0.184 \\
w/o Audio (V+T) & 0.540 & 0.295 & 0.200 \\
w/o Text (V+A)  & 0.538 & 0.293 & 0.198 \\
\midrule
Video only (V)  & 0.485 & 0.255 & 0.175 \\
Audio only (A)  & 0.455 & 0.235 & 0.165 \\
Text only (T)   & 0.440 & 0.228 & 0.160 \\
\bottomrule
\end{tabular}
\vskip -0.1in
\end{table}

\subsection{Frozen Stimulus Encoder Comparison}
\label{sec:appx_encoder_ablation}

Table~\ref{tab:frozen_encoder_ablation} evaluates the effect of replacing the frozen stimulus encoders while keeping the forecasting architecture fixed on Algonauts~2025 at $H=20$.
The main encoder setting performs best, suggesting that temporally specific stimulus representations are important for conditional rollout.
The performance drop under alternative encoders also indicates that BrainVista's gains are not solely due to the autoregressive forecasting architecture, but also depend on the quality and temporal alignment of the stimulus representation.

\begin{table}[t]
\centering
\setlength{\tabcolsep}{4.2pt}
\renewcommand{\arraystretch}{1.05}
\footnotesize
\caption{\textbf{Frozen stimulus encoder ablation on Algonauts~2025 at $H=20$.}
The forecasting architecture is fixed, and only the frozen stimulus encoders are changed.}
\label{tab:frozen_encoder_ablation}
\begin{tabular}{@{}lccc@{}}
\toprule
\textbf{Frozen stimulus encoder(s)} & $r$ & $p_{\mathrm{corr}}$ & Cls@Top10 \\
\midrule
Main paper encoders & \textbf{0.2183} & \textbf{0.2090} & \textbf{0.7619} \\
ImageBind           & 0.2110 & 0.1990 & 0.7420 \\
Qwen Omni 2.5       & 0.2020 & 0.1880 & 0.7240 \\
\bottomrule
\end{tabular}
\vskip -0.1in
\end{table}

\subsection{Cross-subject Generalizability}
\label{sec:appx_cross_subject}

Cross-subject transfer evaluates whether the learned forecasting rule generalizes across individuals.
We conduct this analysis on Algonauts~2025, where all subjects share the same Schaefer-1000 parcel space and Yeo-7 network assignment.
We compare BrainVista with BrainHarmony-AR, the strongest matched-access baseline in the main results.
For single-source cross-subject transfer, the model is trained on one source subject and evaluated on a different target subject without target-subject fine-tuning.
For leave-one-subject-out transfer, the model is trained on three subjects and evaluated on the held-out subject.

\begin{table}[t]
\centering
\setlength{\tabcolsep}{4.4pt}
\renewcommand{\arraystretch}{1.08}
\footnotesize
\caption{\textbf{Cross-subject transfer summary on Algonauts~2025 under $H{=}20$ forecasting.}
Each entry reports subject-level $p_{\mathrm{corr}}$ mean $\pm$ standard deviation.
BH-AR denotes BrainHarmony-AR.}
\label{tab:cross_subject_transfer_summary}
\begin{tabular}{@{}lccc@{}}
\toprule
\textbf{Training / evaluation setting}
& \textbf{BH-AR}
& \textbf{BrainVista}
& $\boldsymbol{\Delta}$ \\
\midrule
Within-subject training
& $0.1887\pm0.0064$
& $\mathbf{0.2090\pm0.0098}$
& $+0.0203$ \\
Single-source cross-subject
& $0.1583\pm0.0108$
& $\mathbf{0.1802\pm0.0119}$
& $+0.0219$ \\
Leave-one-subject-out
& $0.1698\pm0.0106$
& $\mathbf{0.1912\pm0.0118}$
& $+0.0214$ \\
\bottomrule
\end{tabular}
\vskip -0.1in
\end{table}

Table~\ref{tab:cross_subject_transfer_summary} shows that BrainVista retains a positive margin over BrainHarmony-AR under both single-source transfer and leave-one-subject-out transfer.
However, cross-subject performance remains lower than within-subject training, indicating that atlas-level anatomical alignment does not fully remove inter-subject functional variability.
This suggests that future work could combine BrainVista with hyperalignment, subject-specific functional alignment, or lightweight subject-adaptation layers.

\begin{table}[t]
\centering
\setlength{\tabcolsep}{5.2pt}
\renewcommand{\arraystretch}{1.08}
\footnotesize
\caption{\textbf{BrainHarmony-AR cross-subject transfer matrix on Algonauts~2025.}
Columns index the training subject, rows index the evaluation subject, and each entry reports $p_{\mathrm{corr}}$.
Diagonal entries correspond to within-subject training.}
\label{tab:cross_subject_bhar_matrix}
\begin{tabular}{@{}lcccc@{}}
\toprule
\multirow{2}{*}{\textbf{Test subject}}
& \multicolumn{4}{c}{\textbf{Training subject}} \\
\cmidrule(lr){2-5}
& Sub-01 & Sub-02 & Sub-03 & Sub-05 \\
\midrule
Sub-01 & 0.1814 & 0.1628 & 0.1660 & 0.1542 \\
Sub-02 & 0.1605 & 0.1920 & 0.1640 & 0.1558 \\
Sub-03 & 0.1650 & 0.1586 & 0.1856 & 0.1516 \\
Sub-05 & 0.1533 & 0.1560 & 0.1518 & 0.1958 \\
\bottomrule
\end{tabular}
\vskip -0.1in
\end{table}

\begin{table}[t]
\centering
\setlength{\tabcolsep}{5.2pt}
\renewcommand{\arraystretch}{1.08}
\footnotesize
\caption{\textbf{BrainVista cross-subject transfer matrix on Algonauts~2025.}
Columns index the training subject, rows index the evaluation subject, and each entry reports $p_{\mathrm{corr}}$.
Diagonal entries correspond to within-subject training.}
\label{tab:cross_subject_brainvista_matrix}
\begin{tabular}{@{}lcccc@{}}
\toprule
\multirow{2}{*}{\textbf{Test subject}}
& \multicolumn{4}{c}{\textbf{Training subject}} \\
\cmidrule(lr){2-5}
& Sub-01 & Sub-02 & Sub-03 & Sub-05 \\
\midrule
Sub-01 & 0.1970 & 0.1850 & 0.1881 & 0.1717 \\
Sub-02 & 0.1824 & 0.2109 & 0.1859 & 0.1761 \\
Sub-03 & 0.1875 & 0.1813 & 0.2074 & 0.1738 \\
Sub-05 & 0.1762 & 0.1790 & 0.1758 & 0.2207 \\
\bottomrule
\end{tabular}
\vskip -0.1in
\end{table}

The source--target matrices in Tables~\ref{tab:cross_subject_bhar_matrix} and~\ref{tab:cross_subject_brainvista_matrix} show that transfer performance varies across subject pairs.
The diagonal entries are highest, as expected, because within-subject training avoids inter-subject functional misalignment.
Across off-diagonal entries, BrainVista consistently improves over BrainHarmony-AR, suggesting that the proposed network-wise tokenization and temporally valid conditioning are beneficial even when the source and target subjects differ.
Nevertheless, the reduction from within-subject to cross-subject performance highlights a remaining limitation of using a shared atlas-level representation alone.

\FloatBarrier
\section{Model-based Neuroscience Analyses}
\label{sec:appx_model_based_analysis}

\subsection{Network-level Rollout Fidelity}
\label{sec:appx_yeo7_breakdown}

Table~\ref{tab:yeo7_pcorr_h10_subjects} reports Yeo-7 network-averaged $p_{\mathrm{corr}}$ under $H=10$ rollout.
Visual and default-mode systems show higher fidelity, whereas limbic regions remain more difficult to predict.
This pattern is consistent across subjects and suggests that prediction difficulty varies systematically across functional systems.

\begin{table}[t]
\centering
\setlength{\tabcolsep}{4.2pt}
\renewcommand{\arraystretch}{1.05}
\footnotesize
\caption{\textbf{Yeo-7 network-level $p_{\mathrm{corr}}$ under $H=10$ rollout.}}
\label{tab:yeo7_pcorr_h10_subjects}
\begin{tabular}{@{}lccccccc@{}}
\toprule
\textbf{Subject} & Visual & SomMot & DorsAttn & VentrAttn & Limbic & FPN/Control & Default \\
\midrule
sub01 & 0.3512 & 0.2830 & 0.3131 & 0.2673 & 0.1807 & 0.2690 & 0.3237 \\
sub02 & 0.3381 & 0.2605 & 0.2977 & 0.2505 & 0.1695 & 0.2595 & 0.3092 \\
sub03 & 0.3420 & 0.2726 & 0.3020 & 0.2581 & 0.1715 & 0.2592 & 0.3127 \\
sub05 & 0.3275 & 0.2520 & 0.2835 & 0.2415 & 0.1585 & 0.2465 & 0.3015 \\
\bottomrule
\end{tabular}
\end{table}

\subsection{History-token Occlusion}
\label{sec:appx_history_occlusion}

To estimate temporal reliance, we mask the history tokens corresponding to a specific lag while keeping all other inputs unchanged.
We then measure the increase in prediction loss relative to the unoccluded setting.
Larger loss increases indicate stronger reliance on that lagged context.
This analysis is a model-based diagnostic, not a causal neuroscience intervention.
It is intended to measure how the trained model distributes temporal dependence across the available context window.

\subsection{Spatial Mixer Attention}
\label{sec:appx_spatial_attention}

We extract attention weights from the Spatial Mixer Head and aggregate them over time and heads to obtain a compact network-to-network interaction summary.
The resulting attention patterns should be interpreted as model-based summaries of learned cross-network interaction structure, not as direct biological causality.
We therefore describe the observed pattern as \emph{heterogeneous and selective} rather than as a binary statement that some networks are coupled and others are uncoupled.

\subsection{Additional Subject-level Visualizations}
\label{sec:appx_subject_viz}

\begin{figure}[t]
\centering
\includegraphics[width=0.95\columnwidth]{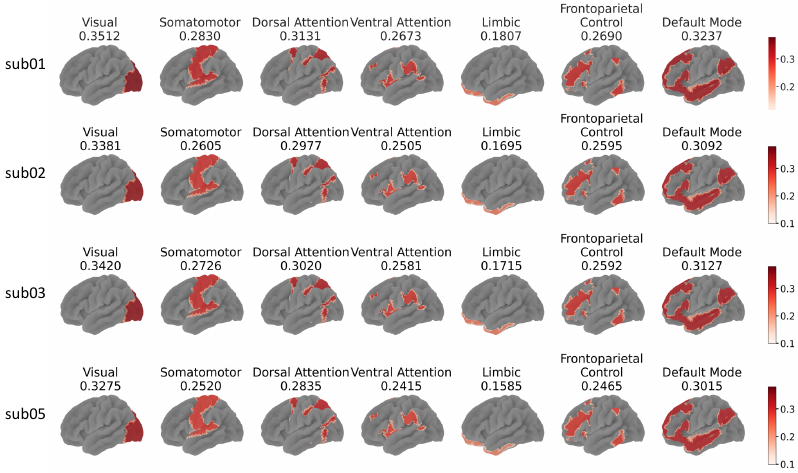}
\caption{\textbf{Additional subject-level model diagnostics on Algonauts 2025.}
We visualize parcel-wise rollout fidelity and Yeo-7 network-level aggregation under conditional autoregressive rollout. The results show spatially structured prediction quality and stable differences across functional systems.}
\label{fig:appx_subject_viz_1}
\end{figure}

\begin{figure}[t]
\centering
\includegraphics[width=0.95\columnwidth]{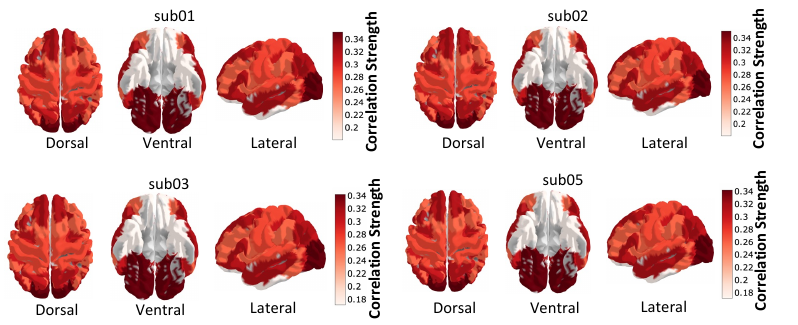}
\caption{\textbf{Additional temporal-reliance diagnostics on Algonauts 2025.}
History-token occlusion measures the increase in prediction loss after masking lagged brain-state tokens. The results show that recent context is important across systems, while higher-order systems retain broader sensitivity to longer temporal histories.}
\label{fig:appx_subject_viz_2}
\end{figure}

\begin{figure}[t]
\centering
\includegraphics[width=0.95\columnwidth]{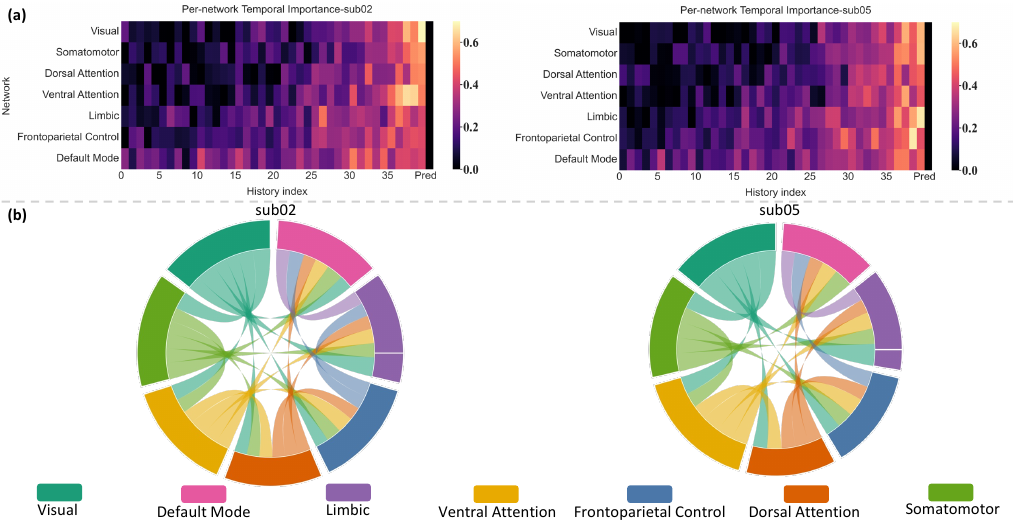}
\caption{\textbf{Additional Spatial Mixer attention diagnostics on Algonauts 2025.}
Spatial Mixer attention is aggregated over time and heads to summarize model-based cross-network interaction patterns. The chord diagrams indicate heterogeneous and selective network interactions rather than uniform global mixing.}
\label{fig:appx_subject_viz_3}
\end{figure}

Figure~\ref{fig:appx_subject_viz_1}--Figure~\ref{fig:appx_subject_viz_3} provide additional subject-level diagnostics on Algonauts 2025.
These visualizations extend the main-text analysis by applying the same diagnostic pipeline to additional subjects and diagnostic views.
The goal is to verify that the temporal-reliance and cross-network interaction patterns discussed in the main paper are not driven by a single favorable subject.

\paragraph{Parcel-wise Rollout Fidelity.}
For each subject, we compute parcel-wise rollout fidelity by correlating the predicted and ground-truth trajectories at each parcel under conditional autoregressive rollout.
The resulting maps show that forecasting quality is spatially heterogeneous rather than uniformly distributed across cortex.
Sensory and default-mode regions generally show higher fidelity, while limbic and some association regions are harder to predict.
This pattern is consistent with the network-level results in Table~\ref{tab:yeo7_pcorr_h10_subjects}.

\paragraph{Network-level Aggregation.}
We aggregate parcel-wise scores to Yeo-7 functional systems to obtain a compact summary of system-level predictability.
Across subjects, visual and default-mode networks tend to preserve higher $p_{\mathrm{corr}}$, whereas limbic regions remain consistently lower.
This supports the interpretation that BrainVista captures stable system-level differences in forecasting difficulty, rather than only improving global average scores.

\paragraph{History-token Occlusion.}
For each lag in the context window, we mask the corresponding history token and measure the increase in prediction loss relative to the unoccluded setting.
The resulting heatmaps show that the model relies strongly on recent history, while several association networks retain broader sensitivity over longer temporal lags.
This analysis is diagnostic only: it measures model reliance on historical tokens and should not be interpreted as direct evidence of biological causality.

\paragraph{Spatial Mixer Attention.}
We further visualize the Spatial Mixer attention after aggregating attention weights over time and heads.
The resulting chord diagrams show heterogeneous and selective cross-network interaction patterns.
The analysis does not imply that the learned attention weights are biological connectivity estimates.
Instead, they provide a model-based summary of which network pairs are emphasized by the forecasting model when producing future brain-region tokens.

Overall, the additional visualizations show that the main qualitative findings persist across subjects: rollout fidelity is spatially structured, temporal reliance is not uniform across functional systems, and the Spatial Mixer learns non-uniform cross-network interactions.
These diagnostics support the claim that BrainVista improves forecasting while preserving interpretable system-level structure.

\FloatBarrier
\section{Downstream Utility: Emotion-category Structure}
\label{sec:appx_downstream_utility}

To evaluate whether predicted brain features preserve behaviorally meaningful structure beyond trajectory-level metrics, we conduct an auxiliary emotion-category clustering analysis on Algonauts~2025 held-out movie segments.
This analysis is intended as a representational diagnostic rather than a clinical, demographic, or individual-level prediction task.

\paragraph{Protocol.}
We divide held-out movie runs into non-overlapping 10-second clips.
For each clip, we obtain an emotion-category label from the corresponding multimodal stimulus content using the same external emotion-labeling pipeline for all methods.
We retain nine emotion categories with sufficient clip counts: happy, sad, neutral, angry, worried, surprise, fear, contempt, and doubt.
Clips with low-confidence or ambiguous labels are excluded from this auxiliary analysis.
For each method, predicted brain features are averaged over the fMRI frames corresponding to each 10-second clip.
The resulting clip-level predicted brain vector is reduced with PCA before clustering.
Unless otherwise specified, we retain 50 principal components.

\paragraph{Clustering Metrics.}
We apply K-means clustering with $K=9$, matching the number of retained emotion categories.
K-means is run with 10 random initializations, and the best solution by within-cluster inertia is used.
We evaluate cluster--label agreement using adjusted Rand index (ARI), normalized mutual information (NMI), and per-class purity.
ARI and NMI measure global agreement between clusters and emotion labels, while per-class purity measures whether clips from each emotion category concentrate in a dominant cluster.

\begin{table}[t]
\centering
\setlength{\tabcolsep}{6.5pt}
\renewcommand{\arraystretch}{1.08}
\footnotesize
\caption{\textbf{Emotion-category clustering from predicted brain features on Algonauts~2025.}
Higher ARI and NMI indicate better agreement between predicted brain-feature clusters and stimulus-derived emotion categories.}
\label{tab:emotion_clustering}
\begin{tabular}{@{}lcc@{}}
\toprule
\textbf{Feature} & ARI & NMI \\
\midrule
Ground-truth brain feature & 0.318 & 0.392 \\
BrainHarmony-AR & 0.138 & 0.214 \\
BrainVista & \textbf{0.214} & \textbf{0.286} \\
\bottomrule
\end{tabular}
\vskip -0.1in
\end{table}

\begin{table}[t]
\centering
\setlength{\tabcolsep}{5.0pt}
\renewcommand{\arraystretch}{1.08}
\footnotesize
\caption{\textbf{Per-class purity in emotion-category clustering on Algonauts~2025.}
Purity is computed between K-means clusters from predicted brain features and stimulus-derived emotion categories.}
\label{tab:emotion_purity}
\begin{tabular}{@{}lcc@{}}
\toprule
\textbf{Emotion category} & BrainHarmony-AR & BrainVista \\
\midrule
Happy    & 0.46 & \textbf{0.58} \\
Sad      & 0.39 & \textbf{0.50} \\
Neutral  & 0.58 & \textbf{0.67} \\
Angry    & 0.42 & \textbf{0.54} \\
Worried  & 0.34 & \textbf{0.44} \\
Surprise & 0.39 & \textbf{0.49} \\
Fear     & 0.36 & \textbf{0.46} \\
Contempt & 0.26 & \textbf{0.35} \\
Doubt    & 0.29 & \textbf{0.38} \\
\midrule
Mean     & 0.39 & \textbf{0.49} \\
\bottomrule
\end{tabular}
\vskip -0.1in
\end{table}

\begin{figure}[t]
\centering
\includegraphics[width=0.95\columnwidth]{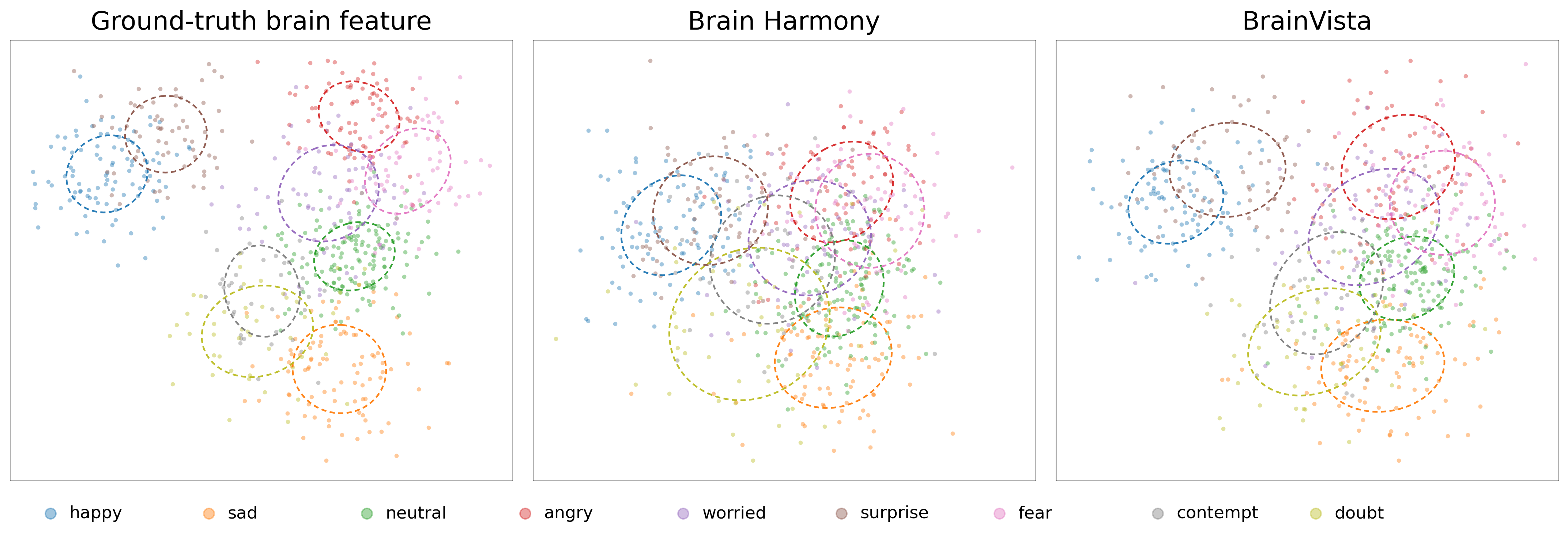}
\caption{\textbf{Emotion-category structure from predicted brain features on Algonauts~2025.}
Predicted brain features are averaged over 10-second clips, reduced with PCA, and clustered with K-means.
Compared with BrainHarmony-AR, BrainVista produces clusters that better align with stimulus-derived emotion-category labels, indicating stronger preservation of affect-related representational structure.}
\label{fig:emotion_clustering_viz}
\end{figure}

BrainVista improves ARI, NMI, and mean per-class purity over BrainHarmony-AR, suggesting that its predicted brain features preserve more stimulus-related affective structure.
The ground-truth brain-feature upper reference remains higher, indicating that the clustering analysis is not saturated and that additional information is still lost during forecasting.
We therefore treat this result as an auxiliary analysis of representational utility rather than as a primary evaluation objective.

\paragraph{Relation to Downstream Prediction.}
Some downstream tasks, such as demographic prediction, disease diagnosis, or clinical progression modeling, would require subject-level labels that are not consistently available in the naturalistic multimodal fMRI datasets used here.
We therefore do not claim clinical, demographic, or individual-level predictive validity.
The emotion-category clustering analysis instead provides a lightweight representational utility test, evaluating whether predicted brain features preserve affect-related structure associated with naturalistic stimuli.

\FloatBarrier
\section{Data and Model Scaling}
\label{sec:appx_scaling}

We study scaling on Algonauts~2025 with fixed pretrained Network-wise Tokenizer--detokenizer pairs, frozen stimulus encoders, validation and test splits, optimization protocol, and rollout evaluation. The purpose is to distinguish additional training data from increased temporal-predictor capacity. The Base model is the configuration used in the main experiments.

\subsection{Training-data Scaling}
\label{sec:appx_data_scaling}

We subsample complete training units, preserving run boundaries rather than sampling overlapping windows across split boundaries. A unit is a complete run or episode used by the training pipeline; an anchor is a valid within-unit context--target position as defined in Appendix~\ref{sec:appx_dataset_statistics}. The retained unit and anchor counts per subject are given in Table~\ref{tab:data_scaling}. Validation and test sets remain fixed across fractions.

\begin{table}[ht]
\centering
\small
\setlength{\tabcolsep}{6pt}
\caption{\textbf{Training-data scaling on Algonauts~2025.} Forecasting columns report $p_{\mathrm{corr}}$. Relative performance is the $H=20$ score divided by the full-data score. Anchor counts are rounded.}
\label{tab:data_scaling}
\begin{tabular}{@{}lrrrrrr@{}}
\toprule
Data & Units/subject & Anchors/subject & $H=1$ & $H=10$ & $H=20$ & Relative \\
\midrule
25\% & 7 & 2.9k & 0.5174 & 0.2589 & 0.1698 & 81.2\% \\
50\% & 14 & 5.9k & 0.5408 & 0.2818 & 0.1874 & 89.7\% \\
75\% & 21 & 8.8k & 0.5528 & 0.2978 & 0.2003 & 95.8\% \\
100\% & 28 & 11.8k & \textbf{0.5600} & \textbf{0.3100} & \textbf{0.2090} & 100\% \\
\bottomrule
\end{tabular}
\end{table}

Performance increases monotonically across the four data fractions at all three horizons. At $H=20$, $p_{\mathrm{corr}}$ increases from $0.1698$ to $0.2090$. This is evidence for a positive data trend over the tested range, not an estimate of a power-law exponent.

\subsection{Model Capacity and Training--Validation Gap}
\label{sec:appx_model_scaling}

Table~\ref{tab:model_scaling} compares Small, Base, and Large configurations. Only the predictor capacity is varied; the pretrained tokenizers and stimulus encoders are fixed.

\begin{table}[ht]
\centering
\small
\setlength{\tabcolsep}{7pt}
\caption{\textbf{Model-capacity scaling on Algonauts~2025.} Scores are $p_{\mathrm{corr}}$. Compute is reported in GPU-hours per subject.}
\label{tab:model_scaling}
\begin{tabular}{@{}lcccc@{}}
\toprule
Model & GPU-h/subject & $H=1$ & $H=10$ & $H=20$ \\
\midrule
Small & 1.4 & 0.5417 & 0.2913 & 0.1938 \\
Base & 2.1 & 0.5600 & 0.3100 & 0.2090 \\
Large & 4.9 & \textbf{0.5641} & \textbf{0.3134} & \textbf{0.2118} \\
\bottomrule
\end{tabular}
\end{table}

Increasing capacity from Small to Base improves forecasting. Increasing it further to Large improves $H=20$ by only $0.0028$, while reported computation increases from $2.1$ to $4.9$ GPU-hours per subject. Table~\ref{tab:scaling_generalization_gap} examines training and validation performance at the selected checkpoint.

\begin{table}[ht]
\centering
\small
\setlength{\tabcolsep}{7pt}
\caption{\textbf{Training--validation gaps in the capacity comparison.} Scores are $p_{\mathrm{corr}}$ at $H=20$; the gap is training minus validation performance. Best epoch denotes the selected epoch out of 40.}
\label{tab:scaling_generalization_gap}
\begin{tabular}{@{}lccccc@{}}
\toprule
Model & Train & Validation & Gap & Best epoch/40 & Test \\
\midrule
Small & 0.2178 & 0.1902 & 0.0276 & 38 & 0.1938 \\
Base & 0.2611 & 0.2047 & 0.0564 & 34 & 0.2090 \\
Large & 0.3094 & 0.2069 & 0.1025 & 27 & 0.2118 \\
\bottomrule
\end{tabular}
\end{table}

The Large model fits training data better, but its validation improvement is small, its training--validation gap is larger, and its best epoch occurs earlier. These observations are consistent with diminishing returns and emerging overfitting on this dataset. Four data fractions and three model sizes on one dataset do not establish a general neural scaling law. The results motivate testing additional naturalistic fMRI data before assuming that larger predictors will continue to improve performance.

\FloatBarrier
\section{Computational Cost and Reproducibility}
\label{sec:appx_complexity}

Table~\ref{tab:model_complexity_by_dataset} summarizes model dimensions, training settings, and computational cost.
Because $d_{\mathrm{brain}}$ depends on the model-space brain representation, the brain projection layers are dataset-dependent.

\begin{table}[htbp]
\centering
\setlength{\tabcolsep}{4.0pt}
\renewcommand{\arraystretch}{1.08}
\scriptsize
\caption{\textbf{Model configuration and computational cost by dataset.}
The temporal predictor uses the same hidden dimension across datasets; brain input/output projection dimensions follow the dataset-specific latent representation.}
\label{tab:model_complexity_by_dataset}
\begin{tabular}{@{}lccc@{}}
\toprule
\textbf{Component / metric} & \textbf{Algonauts 2025} & \textbf{CineBrain} & \textbf{HAD} \\
\midrule
Raw brain dimension $\sum_k Q_k$ & 1,000 & 18,946 & fsLR surface \\
Model-space input dimension $\sum_k P_k$ & 1,000 & 1,024 & 1,000 \\
Concatenated brain latent $d_{\mathrm{brain}}$ & 803 & 820 & 803 \\
Transformer hidden size $d_{\mathrm{model}}$ & 256 & 256 & 256 \\
Brain projection & $803\leftrightarrow256$ & $820\leftrightarrow256$ & $803\leftrightarrow256$ \\
\midrule
Peak training memory & 24.0 GB & 26.8 GB & 24.3 GB \\
Inference time for $H=20$ rollout & 0.18 s & 0.20 s & 0.18 s \\
Default context length & $L=40$ & $L=40$ & $L=40$ \\
Default rollout horizon & $H=20$ & $H=20$ & $H=20$ \\
Optimizer & AdamW & AdamW & AdamW \\
Learning rate & $2\times10^{-4}$ & $2\times10^{-4}$ & $2\times10^{-4}$ \\
Weight decay & $10^{-2}$ & $10^{-2}$ & $10^{-2}$ \\
Batch size & 32 & 32 & 32 \\
Training epochs & 40 & 40 & 40 \\
Mixed precision & PyTorch AMP & PyTorch AMP & PyTorch AMP \\
\bottomrule
\end{tabular}
\end{table}

All experiments were conducted on a single NVIDIA GeForce RTX 5090 GPU using PyTorch automatic mixed precision.
The reported wall-clock training time includes both network-wise tokenizer pretraining and temporal predictor training.
Table~\ref{tab:compute_training_time} summarizes the compute resources used for each dataset.
All models fit within a single-GPU training setup without model parallelism or distributed training.

\FloatBarrier
\section{Limitations and Assumptions}
\label{sec:appx_limitations}

BrainVista assumes that the exogenous stimulus stream is observed during forecasting.
Therefore, the task is conditional brain-state forecasting rather than unconditional brain simulation.
The model also assumes that past fMRI context is available for rollout initialization.
Forecasting accuracy decreases as the horizon grows because prediction errors accumulate recursively.
In addition, association systems remain harder to predict than sensory systems, and the present analyses do not isolate the causes of this difference.

The Spatial Mixer Head provides useful model-based diagnostics of learned cross-network interactions, but its attention weights should not be interpreted as direct biological causality.
All main results use training-set normalization to avoid evaluation-time leakage from validation or test runs.
CineBrain uses training-set PCA to reduce high-dimensional voxel-level responses before tokenization, and the resulting PCA basis is fixed for validation and test runs.
Finally, the three evaluated datasets differ in stimulus protocol, temporal resolution, and preprocessing pipeline.
Models are trained separately on each dataset. These results evaluate held-out runs or episodes under the official splits, not arbitrary unseen movies, and do not establish zero-shot transfer across scanners or protocols. Cross-subject transfer in Appendix~\ref{sec:appx_multimodal_transfer} has lower absolute accuracy than within-subject training.

BrainVista is a within-run predictive assay for comparing dynamical assumptions and stimulus representations against withheld fMRI. It does not establish accurate cross-session trajectory simulation, pre-scan stimulus evaluation, online experimental control, clinical diagnosis, or individual-level decision-making validity. History occlusion measures conditional predictive reliance, not a biological intervention; attention weights do not establish direct connectivity. The emotion analysis uses stimulus-derived labels and is an auxiliary representational diagnostic, not a measure of participants' subjective emotional states. Neuroimaging data remain subject to their original access and privacy conditions; forecasted responses should not be used to infer sensitive individual attributes.

\clearpage

\end{document}